\documentclass[11pt]{article}
\usepackage{amsfonts}
\usepackage{amssymb}
\usepackage{amsmath,amssymb}
\usepackage{bbm}

\def\slasha#1{\setbox0=\hbox{$#1$}#1\hskip-\wd0\hbox to\wd0{\hss\sl/\/\hss}}

\def\periodb#1{\setbox0=\hbox{$#1$}#1\hskip-\wd0\hbox to\wd0{-}}

\newcommand{\pZ}{p}    
\def\sfrac#1#2{{\textstyle\frac{#1}{#2}}}
\newcommand{\unit}{\mathbbm{1}}   
\newcommand{\frC}{\mathfrak{C}}
\newcommand{\CN}{\mathcal{N}}    
\newcommand{\CA}{\mathcal{A}}    
 
\newcommand{\CB}{\mathcal{B}} 
\newcommand{\CD}{\mathcal{D}} 
\newcommand{\CO}{\mathcal{O}}    
\newcommand{\CV}{\mathcal{V}}    
\newcommand{\CL}{\mathcal{L}}    
\newcommand{\CU}{\mathcal{U}}    
\newcommand{\CF}{\mathcal{F}}    
\newcommand{\CP}{\mathcal{P}}    
\newcommand{\CT}{\mathcal{T}}    
\newcommand{\CE}{\mathcal{E}}    
\newcommand{\CY}{\mathcal{Y}}  
\newcommand{\FZ}{\mathbbm{Z}}     
\newcommand{\FR}{\mathbbm{R}}     
\newcommand{\FC}{\mathbbm{C}}     
\newcommand{\CPP}{{\mathbbm{C}P}}    
\newcommand{\RZ}{\mathbbm{Z}}     
\newcommand{\dd}{\mathrm{d}}     
\newcommand{\dpar}{\partial}     
\newcommand{\hra}{{\hookrightarrow}}     
\newcommand{\diag}{{\mathrm{diag}}}     
\newcommand{\cb}{\,,\,}             
\newcommand{\di}{\mathrm{i}}     
\newcommand{\bl}{{\bar{\lambda}}}     
\newcommand{\ald}{{\dot{\alpha}}}     
\newcommand{\bed}{{\dot{\beta}}}     
\newcommand{\hx}{{\hat{x}}}  

\newcommand{\al}{{{\alpha}}}     
\newcommand{\vk}{{{\varkappa}}}

\makeatletter
\renewcommand*\l@section{\@dottedtocline{1}{1.5em}{4em}}
\@addtoreset{equation}{section}
\makeatother
\renewcommand{\thesection}{\arabic{section}.}
\renewcommand{\theequation}{\thesection\arabic{equation}}

\begin{document}
\begin{titlepage}
\setcounter{page}{0}
.
\vskip 3.5cm
\begin{center}
{\LARGE \bf On Exact Solvability of $\CN$=4 super Yang-Mills}
\vskip 1.5cm
{\Large Alexander D. Popov}
\vskip 1cm
{\em Institut f\"{u}r Theoretische Physik,
Leibniz Universit\"{a}t Hannover\\
Appelstra{\ss}e 2, 30167 Hannover, Germany}\\
{Email: alexander.popov@itp.uni-hannover.de}
\vskip 1.1cm
\end{center}
\begin{center}
{\bf Abstract}
\end{center}
We consider the ambitwistor description of $\CN$=4 supersymmetric extension of U($N$) Yang-Mills theory on Minkowski space
$\FR^{3,1}$. It is shown that solutions of super-Yang-Mills equations are encoded in real analytic U($N$)-valued functions on a
domain in superambitwistor space $\CL_{\FR}^{5|6}$ of real dimension $(5|6)$. This leads to a procedure for generating 
solutions of super-Yang-Mills equations on $\FR^{3,1}$ via solving a Riemann-Hilbert-type factorization problem on two-spheres in
 $\CL_{\FR}^{5|6}$.

\end{titlepage}
\newpage
\setcounter{page}{1}

\section{Introduction and summary}

\noindent
There are many indications that $\CN$=4 supersymmetric Yang-Mills (SYM) theory in four dimensions might be
integrable (see e.g. reviews~\cite{1}-\cite{6} and references therein). Maximally supersymmetric U($N$) gauge theory is the simplest 
interacting conformal field theory in four dimensions, which is a useful model, the study of which can help to understand more
realistic gauge field theories. In the planar $N\to\infty$ limit the quantum $\CN$=4 SYM model exhibits integrability, which manifests
itself through an infinite-dimensional extension of the superconformal symmetry algebra $\mathfrak{psu}(2, 2|4)$ to the Yangian 
Y$[\mathfrak{psu}(2, 2|4)]$
(see e.g.~\cite{1}-\cite{8} and references therein). Despite many successes related to the planar integrability of $\CN$=4 
SYM theory, the situation is far from satisfactory, as was discussed e.g. in~\cite{9, 10}. In fact, only the quantum planar limit $N\to\infty$
was considered, selecting special quantum observables. However, it is not clear what is the picture for U($N$) SYM at finite $N$ and
what can one say about integrability of  $\CN$=4 SYM on the classical level. These are exactly the issues that we want to discuss 
in this paper. 

Integrability of the self-dual subsector of classical $\CN \le 4$ SYM theories was studied by Wolf~\cite{11}. He described the twistor
correspondence between solutions of $\CN$-extended self-dual SYM equations and transition functions of holomorphic vector bundles 
over the supertwistor space $\CP^{3|\CN}:=\CPP^{3|\CN}{\setminus}\CPP^{1|\CN}$ and found, in particular, affine extensions of the
gauge algebra $\mathfrak{u}(N)$ and the superconformal algebra as infinitesimal hidden symmetries acting on the solution space of self-dual SYM.
Hidden symmetries of the full $\CN$=4 SYM model were also considered~\cite{12}, but with much less generality -- only an affine extension
of the algebra of supertranslations was found. 

In this paper, we consider the ambitwistor description of  $\CN$=4 SYM theory and show that this theory is solvable on the classical level. 
To achieve this goal, we introduce the twistor space $\CP^{3}:=\CPP^{3}{\setminus}\CPP^{1}$ as an open subset of complex projective 
space $\CPP^{3}$ and an $\CN$-extended supertwistor space $\CP^{3|\CN}$ associated with the complex super Minkowski space $\FC^{4|4\CN}$.
Then the complex {\it superambitwistor} space $\CL_{\FC}^{5|6}$ is defined as a quadric hypersurface in the direct product 
$\CP^{3|3}\times\CP^{3|3}_\ast$ of two supertwistor spaces with $\CN$=3.  The space $\CL_{\FC}^{5|6}$ parametrizes complex 
$(1|6)$-dimensional super null lines in super Minkowski space $\FC^{4|12}$. On the other hand, points $\hx$ in $\FC^{4|12}$
correspond to subspaces $Q^2_\hx\cong\CPP^1\times\CPP^1_\ast$ in $\CL_{\FC}^{5|6}$. There is a well-known one-to-one
correspondence between gauge equivalence classes of solutions to the $\CN$=3 SYM equations on complexified Minkowski space $\FC^4$ and
equivalence classes of holomorphic vector bundles $\CE$ over $\CL_{\FC}^{5|6}$ holomorphically trivial on each submanifold 
$Q^2_\hx\hra\CL_{\FC}^{5|6}$ with $\hx\in\FC^{4|12}$~\cite{13}-\cite{18}. Recall that $\CN$=3 and $\CN$=4 SYM theories have the 
same physical content, i.e. they coincide on shell. Using a proper reality condition, we introduce a real superambitwistor space 
$\CL_{\FR}^{5|6}\subset\CL_{\FC}^{5|6}$ which is covered by two patches $\CU_\pm$, $\CL_{\FR}^{5|6}=\CU_+\cup\CU_-$. Then
real analytic solutions of $\CN =3$ SYM model on Minkowski space $\FR^{3,1}$ correspond to free U($N$)-valued real analytic functions $f_{+-}$
defined on $\CU_+\cap\CU_-\subset\CL_{\FR}^{5|6}$. Solutions of  Riemann-Hilbert-type factorization problems for such $f_{+-}$ 
restricted to 2-spheres $S^2_\hx$ in $\CL_{\FR}^{5|6}$ give real analytic solutions of $\CN$=3 SYM equations on $\FR^{3,1}$ and hence 
solvability of the model.

By solvability we mean a description of solutions of differential equations in terms of unconstrained data, and in this sense $\CN$=3 SYM theory
on complexified Minkowski space  $\FC^4$ is not solvable or integrable. The reason is that solutions of the $\CN$=3 SYM equations correspond
to a collection $\{f_{ab}\}$ of transition matrices in holomorphic bundles $\CE$ over $\CL_{\FC}^{5|6}$, and since  $\CL_{\FC}^{5|6}$ is covered by at least 
four patches $\CU_a$, these matrices must satisfy nonlinear algebraic equations at the intersections $\CU_a\cap\CU_b$ of patches (1-cocycle constraints),
i.e. twistor data are not unconstrained. In the twistor description, one can move from $\{f_{ab}\}$ to a superconnection $A(x, \theta , \eta )$ that encodes the
entire $\CN$=3 SYM multiplet and is defined on the Minkowski superspace $\FC^{4|12}$ with coordinates $x,  \theta^i, \eta_i, i=1,2,3$. The superconnection
satisfies zero-curvature equations which are equivalent to the $\CN$=3 SYM field equations for the SYM multiplet. 

The described twistor correspondence also makes it possible to obtain solutions of the bosonic Yang-Mills (YM) equations, which is described as follows.
Let us introduce the nilpotent variable $\tau =\theta^i\eta_i$ and suppose that $A(x, \theta , \eta )$ depends on the Grassmann variables only through $\tau$.
Then $A(x, \tau)=A(x)+B_k(x)\tau^k, k=1,2,3$, since $\tau^4=0$ and it was shown that zero-curvature equations for $A(x, \tau)$ are satisfied iff 
$A(x)$ satisfies the bosonic YM equations on $\FC^4$~\cite{13}-\cite{18}. This connection between solutions of $\CN$=3 SYM and bosonic YM equations
was proved by the methods of algebraic geometry only for {\it complex analytic} solutions on the complexified Minkowski space.

After imposing the reality conditions, we obtain {\it real analytic} solutions of $\CN$=3 SYM and bosonic YM equations on Minkowski space $\FR^{3,1}$.
Thus, we obtain a narrow class of real analytic solutions on $\FR^{3,1}$ {\it extendable}\footnote{Not every real analytic function defined on the whole 
real line $\FR$ can be extended to a holomorphic function on the whole complex plane $\FC$. The standard example is the function
$f(x)=(1+x^2)^{-1}$, $x\in\FR$, whose Teylor series diverges at $|x|>1$ when $x\in\FC$.}  to holomorphic solutions on $\FC^4$. 
These solutions are described by real analytic bundles over
$\CL_{\FR}^{5|6}$, and since $\CL_{\FR}^{5|6}$ can be covered by two patches $\CU_\pm$, the solutions are described by unconstrained real
analytic transition matrices $f_{+-}$ on $\CU_+\cap\CU_-\subset\CL_{\FR}^{5|6}$. Thus, we show that subspace of real analytic solutions of $\CN$=3 SYM 
(and bosonic YM) equations is described by unconstrained data on the superambitwistor space $\CL_{\FR}^{5|6}$, i.e. the model is solvable. This is not true
for solutions with less smoothness, and therefore it cannot be said that the theory under consideration is completely integrable. Integrability appears only 
on a subsector of real analytic solutions.

\section{Twistors, supertwistors and self-dual Yang-Mills}

\noindent
{\bf Twistors.}  We start with a complex projective space $\CPP^3$ with homogeneous (a.k.a. projective) coordinates $(\omega^\al , \lambda_\ald )$
subject to the identification $(\omega^\al , \lambda_\ald )\sim (\vk \omega^\al , \vk\lambda_\ald )$ for any nonzero complex number $\vk , \ \al =1,2$ and
$\ald=1,2$. This means that $\CPP^3$ is the space of one-dimensional subspaces (lines) in the space $\FC^4$ with coordinates 
$(\omega^\al , \lambda_\ald )$. Consider now the space $\CP^3:=\CPP^{3}{\setminus}\CPP^{1}$ in which $(\lambda_\ald )\ne 0$. This space 
can be covered by two patches $\CU_+\ (\lambda_{\dot 1}\ne 0)$ and $\CU_-\ (\lambda_{\dot 2}\ne 0)$ with coordinates\footnote{We introduce open
covering $\CU =\{\CU_a\}$ of manifolds, patches and local coordinates because we use the \v{C}ech approach to describing vector bundles via transition functions.}
\begin{equation}\label{2.1}
z_+^\alpha=\frac{\omega^\alpha}{\lambda_{\dot{1}}}~,~~~
z_+^3=\frac{\lambda_{\dot{2}}}{\lambda_{\dot{1}}}=:\lambda_+~~~
\mbox{on}~~\CU_+~~~ \mbox{and} ~~~
z_-^\alpha=\frac{\omega^\alpha}{\lambda_{\dot{2}}}~,~~~
z_-^3=\frac{\lambda_{\dot{1}}}{\lambda_{\dot{2}}}=:\lambda_-~~~
\mbox{on}~~\CU_-~,
\end{equation}
related by
\begin{equation}\label{2.2}
z_+^\alpha=z_+^3z_-^\alpha~~~\mbox{ and }~~~z_+^3=\frac{1}{z_-^3}
\end{equation}
on the overlap $\CU_+\cap\CU_-$. From \eqref{2.1} and \eqref{2.2} it follows that
$\CP^3=\CU_+\cup\,\CU_-$ coincides with the total space of the
rank 2 holomorphic vector bundle\footnote{The holomorphic vector
bundle $\CO(n)$ over $\CPP^1$ has the transition function
$f_{+-}=\lambda_+^n$ and the first Chern number $c_1=n$.} 
$\CP^3=\CO(1)\oplus\CO(1)$ over $\CPP^1$,
\begin{equation}\label{2.3}
\CP^3\rightarrow\CPP^1~,
\end{equation}
where the Riemann sphere $\CPP^1$ is covered by two patches,
\begin{equation}\label{2.4}
U_+=\CU_+\cap\CPP^1~~~\mbox{and}~~~U_-=\CU_-\cap\CPP^1
\end{equation}
with coordinates $\lambda_+$ on $U_+$ and $\lambda_-$ on $U_-$,
$\CPP^1=U_+\cup U_-$.

A correspondence between the twistor space $\CP^3$ and the complexified Minkowski space $\FC^4$ can be established as follows. 
Consider holomorphic sections of the complex vector bundle \eqref{2.3} defined by the equations
\begin{equation}\label{2.5}
z_\pm^\alpha=x^{\alpha\dot{\alpha}}\lambda_\ald^\pm
\end{equation}
and parametrized by moduli $x=(x^{\alpha\ald})\in\FC^4$, where
\begin{equation}\label{2.6}
\left(\lambda_\ald^+\right):=\left(\begin{array}{c}1\\\lambda_+
\end{array}\right)~~~\mbox{and}~~~
\left(\lambda_\ald^-\right):=\left(\begin{array}{c}\lambda_-\\1
\end{array}\right)~~~\mbox{for}~~~\lambda_\pm\in U_\pm\ .
\end{equation}
Equations \eqref{2.5} allow us to introduce a double fibrations
\begin{equation}\label{2.7}
\begin{picture}(50,40)
\put(0.0,0.0){\makebox(0,0)[c]{$\CP^3$}}
\put(64.0,0.0){\makebox(0,0)[c]{$\FC^4$}}
\put(34.0,33.0){\makebox(0,0)[c]{$\CF^5$}}
\put(7.0,18.0){\makebox(0,0)[c]{$\pi_2$}}
\put(55.0,18.0){\makebox(0,0)[c]{$\pi_1$}}
\put(25.0,25.0){\vector(-1,-1){18}}
\put(37.0,25.0){\vector(1,-1){18}}
\end{picture}
\end{equation}
where $\CF^5:=\FC^4\times\CPP^1$. From this diagram based on \eqref{2.5}, one observes that a point $x=(x^{\alpha\ald})\in\FC^4$ 
corresponds to the projective line $\CPP^1_x=\pi_2(\pi_1^{-1}(x))$ in $\CP^3$ given by \eqref{2.5} for fixed $x$, and a point
$\pZ=(z^\alpha_\pm,\lambda_\ald^\pm)\in\CP^3$ corresponds to a totally null anti-self-dual 2-plane ($\beta$-plane) $\pi_1(\pi_2^{-1}(\pZ))$ in
$\FC^4$ defined by \eqref{2.5} for fixed $(z^\alpha_\pm,\lambda_\ald^\pm)$ and varied $x\in\FC^4$. The double fibration \eqref{2.7} and its 
$\RZ_2$-graded generalizations play the central role in the (super)twistor correspondence.

\noindent
{\bf Supertwistors.} A super extension of the space $\CPP^3$ is the supermanifold $\CPP^{3|\CN}$ with homogeneous coordinates
$(\omega^\alpha , \lambda_\ald , \eta_i)$ subject to the identification 
$(\omega^\alpha , \lambda_\ald , \eta_i)\sim (\vk \omega^\alpha , \vk\lambda_\ald , \vk\eta_i)$ 
for any nonzero complex number $\vk$. Here $(\omega^\alpha , \lambda_\ald )$ are projective coordinates on $\CPP^3$ 
and $\eta_i$ with $i=1,...,\CN$ are Grassmann variables.

We consider the twistor space $\CP^3=\CPP^3{\setminus}\CPP^1$ and its super extension $\CP^{3|\CN}=\CPP^{3|\CN}{\setminus}\CPP^{1|\CN}$ covered
by two patches $\hat\CU_\pm$, $\CP^{3|\CN}=\hat\CU_+\cup\hat\CU_-$, with even coordinates \eqref{2.1} and odd coordinates
\begin{equation}\label{2.8}
\eta_i^+=\frac{\eta_i}{\lambda_{\dot{1}}}~~\mbox{on}~~\hat{\CU}_+
~~~\mbox{and}~~~\eta_i^-=\frac{\eta_i}{\lambda_{\dot{2}}}~~
\mbox{on}~~\hat{\CU}_-
\end{equation}
related by
\begin{equation}\label{2.9}
\eta^+_i=z_+^3\eta^-_i
\end{equation}
on $\hat{\CU}_+\cap\hat{\CU}_-$. We see from \eqref{2.8} and \eqref{2.9} that the fermionic coordinates are sections
of the bundle $\CO(1)\otimes\Pi\FC^\CN$, where $\Pi$ is the operator inverting parity of coordinates. The supermanifold $\CP^{3|\CN}$ is fibred over $\CPP^1$,
\begin{equation}\label{2.10}
\CP^{3|\CN}\rightarrow\CPP^{1}~, \quad \CP^{3|\CN}=\CO(1){\otimes}\FC^2\oplus \CO(1){\otimes}\Pi\FC^\CN\ ,
\end{equation}
with superspaces $\FC^{2|\CN}_\lambda$ as fibres over $\lambda\in\CPP^{1}=\CPP^{1|0}$. 

Holomorphic sections of the bundle \eqref{2.10} are rational curves $\CPP^{1}_{\hat x}$ , defined by the equations~\cite{19}
\begin{equation}\label{2.11}
z_\pm^\alpha=(x^{\alpha\ald}-\theta^{\alpha i}\eta^\ald_i)
\lambda_\ald^\pm~,~~~\eta_i^\pm=\eta_i^\ald\lambda_\ald^\pm
\end{equation}
and parametrized by supermoduli $\hat x=(x,\theta ,\eta )=(x^{\alpha\ald}, \theta^{\alpha i}, \eta^\ald_i)\in\FC^{4|4\CN}$ (super Minkowski space).
Therefore, we obtain a double fibration
\begin{equation}\label{2.12}
\begin{picture}(50,40)
\put(0.0,0.0){\makebox(0,0)[c]{$\CP^{3|\CN}$}}
\put(74.0,0.0){\makebox(0,0)[c]{$\FC^{4|4\CN}$}}
\put(42.0,37.0){\makebox(0,0)[c]{$\CF^{5|4\CN}$}}
\put(7.0,20.0){\makebox(0,0)[c]{$\pi_2$}}
\put(65.0,20.0){\makebox(0,0)[c]{$\pi_1$}}
\put(25.0,27.0){\vector(-1,-1){18}}
\put(47.0,27.0){\vector(1,-1){18}}
\end{picture}
\end{equation}
\medskip
with coordinates
$(z,\lambda , \eta )$ on $\CP^{3|\CN}$, $(x, \lambda , \theta , \eta )$ on $\CF^{5|4\CN}:=\FC^{4|4\CN}\times\CPP^1$, 
and $(x, \theta , \eta )$ on $\FC^{4|4\CN}$. The double fibration \eqref{2.12} defines the following twistor correspondence: a point
$\hat x = (x, \theta, \eta )\in \FC^{4|4\CN}$ corresponds to the projective line $\CPP^1_{\hat x}=\pi_2(\pi_1^{-1}(\hat x))\hra\CP^{3|\CN}$,
and a point $\hat p\in\CP^{3|\CN}$ corresponds to a totally null $\beta$-superplane $\pi_1(\pi_2^{-1}(\hat p))\hra\FC^{4|4\CN}$ of dimension
$(2|3\CN )$.

\noindent
{\bf Dual supertwistors.} We described supermanifold $\CPP^{3|\CN}$ as the space of $(1|0)$-dimensional subspaces in the space $\FC^{4|\CN}$. Its
{\it dual} supermanifold can be defined as a space of $(3| \CN )$-dimensional planes in $\FC^{4|\CN}$ parametrized by homogeneous coordinates
$(\mu_\alpha , \sigma^\ald , \theta^i)$ subject to the identification
$(\mu_\alpha , \sigma^\ald , \theta^i)\sim (\vk\mu_\alpha , \vk\sigma^\ald , \vk\theta^i)$ for any complex number $\vk\in\FC^\ast$. We again have 
the projective superspace $\CP_\ast^{3|\CN}=\CPP^{3|\CN}{\setminus}\CPP_\ast^{1|\CN}=(\CO(1)\oplus\CO(1))^{3|\CN}=\hat\CV_+\cup\hat\CV_-$
with coordinates
\begin{eqnarray}\label{2.13}
&&w_+^\ald=\frac{\sigma^\ald}{\mu_1}~,~~
w_+^{\dot3}=\frac{\mu_2}{\mu_1}=\mu_+~~ \mbox{and} ~~
\theta_+^i=\frac{\theta^i}{\mu_1}~~ \mbox{on} ~~ \hat{\CV}_+~,\\
\label{2.14}
&&w_-^\ald=\frac{\sigma^\ald}{\mu_2}~,~~
w_-^{\dot3}=\frac{\mu_1}{\mu_2}=\mu_-~~ \mbox{and} ~~
\theta_-^i=\frac{\theta^i}{\mu_2}~~ \mbox{on} ~~ \hat{\CV}_-~,\\
\label{2.15}
&&w_+^\ald=\mu_+w_-^\ald~,~~~\mu_+=\mu_-^{-1}~, ~~~
\theta_+^i=\mu_+\theta^i_-~~~\mbox{on}~~~\hat{\CV}_+\cap\hat{\CV}_-~.
\end{eqnarray}
Here, $\mu_\pm$ are coordinates on the patches ${V}_\pm =\hat{\CV}_\pm\cap\CPP_\ast^1$ covering the base $\CPP_\ast^1$ of the 
holomorphic vector bundle
\begin{equation}\label{2.16}
\CP^{3|\CN}_*\rightarrow\CPP^{1}_*~,
\end{equation}
having the same structure as the bundle \eqref{2.10}.

Sections of the bundle \eqref{2.16} (curves $\CPP_\ast^1$ in $\CP^{3|\CN}_*$) are defined by the equations~\cite{19}
\begin{equation}\label{2.17}
w_\pm^\ald=(x^{\alpha\ald}+\theta^{\alpha
i}\eta_i^\ald)\mu_\alpha^\pm~~~\mbox{and}~~~ \theta^i_\pm
=\theta^{\alpha i}\mu_\alpha^\pm\ .
\end{equation}
For the dual supertwistors $\CP^{3|\CN}_*$ we can introduce a double fibration similar to \eqref{2.12} with $\CP^{3|\CN}_*$ instead of $\CP^{3|\CN}$
and $\CF^{5|4\CN}_*=\FC^{4|4\CN}\times\CPP_\ast^1$ instead of $\CF^{5|4\CN}$. Via the equations \eqref{2.17} a point $\hat p=(w,\mu ,\theta )$ in 
$\CP^{3|\CN}_*$ corresponds to a totally null $\alpha$-superplane in $\FC^{4|4\CN}$ with dimension $(2|3\CN)$.

\noindent
{\bf  (Anti-)self-dual (super-)Yang-Mills.} The importance of the twistor diagram \eqref{2.7} lies in the fact that there is a one-to-one correspondence
between gauge equivalence classes of holomorphic solutions to the self-dual Yang-Mills equations on $\FC^4$ and equivalence classes of holomorphic
vector bundles $\CE$ over $\CP^3$ trivial on $\CPP_x^1\hra\CP^3$, $x\in\FC^4$ \cite{20}-\cite{22}.  The map between vector bundles $\CE\to\CP^3$ and
(trivial) vector bundles $E\to\FC^4$ with a self-dual connection is called the Penrose-Ward transform~\cite{22}. This map and its inverse are the essence of the 
twistor approach. The twistor correspondence can be generalized to $\CN$-extended supersymmetric self-dual Yang-Mills theory \cite{23}-\cite{25} by considering the
diagram \eqref{2.12} instead of \eqref{2.7} \cite{11,26,27}. That is, there is a bijection between the moduli space of holomorphic bundles $\CE$ over $\CP^{3|\CN}$
and the moduli space of complex analytic solutions of self-dual $\CN$-extended SYM theory on $\FC^4$. A one-to-one correspondence between gauge
equivalence classes of holomorphic solutions to the $\CN$-extended anti-self-dual Yang-Mills equations on $\FC^4$ and equivalence classes of holomorphic
vector bundles $\CE$ over the dual supertwistor space $\CP^{3|\CN}_\ast$ is described by the double fibration with $\CF^{5|4\CN}_\ast$ and $\CP^{3|\CN}_\ast$
instead of $\CF^{5|4\CN}$ and $\CP^{3|\CN}$ in the self-dual case.

\noindent
{\bf  Integrability.} Note that the diagram \eqref{2.7} can be considered as a particular case of the diagram \eqref{2.12} with $\CN =0$ and one can discuss
\eqref{2.12} for any $\CN\ge 0$ including the bosonic case. The supertwistor space $\CP^{3|\CN}$ in \eqref{2.12} can be covered by two patches $\hat\CU_\pm$
from \eqref{2.8}-\eqref{2.9}. Hence, any holomorphic vector bundle $\CE$ over $\CP^{3|\CN}$ is defined by {\it one} GL$(N, \FC)$-valued function $f_{+-}$
on $\hat\CU_+\cap\hat\CU_-$ (a transition function of the bundle $\CE\to\CP^{3|\CN}$) and after restriction of $\CE$ to $\CPP_{\hat x}^1\hra\CP^{3|\CN}$
we obtain a nonsingular holomorphic matrix-valued function $f_{+-}(\hat x, \lambda )$ on $U_+\cap U_-\subset \CPP_{\hat x}^1$. Then a {\it parametric
Riemann-Hilbert problem} is to find matrix-valued functions $\psi_\pm$ on $U_+\cap U_-$ such that $\psi_+$ can be extended continuously to a regular
GL$(N, \FC)$-valued function on $U_+$, $\psi_-$ can be extended to a regular GL$(N, \FC )$-function on $U_-$ and
\begin{equation}\label{2.18}
f_{+-}(\hat x, \lambda )=\psi_+^{-1}(\hat x, \lambda_+ )\psi_-(\hat x, \lambda_-)
\end{equation}
on $U_+\cap U_-$. Hence, solving $\CN$-extended self-dual Yang-Mills equations on $\FC^4$ (the self-dual subsector of $\CN$-extended super-Yang-Mills) is
reduced to the parametric Riemann-Hilbert problem  \eqref{2.18}, i.e. this model is an integrable system. Infinite hidden symmetries of this self-dual subsector
of SYM theory on $\FC^4$ were described in~\cite{11,27}.

\section{Superambitwistors}

\noindent
{\bf  Quadric $\CL_{\FC}^{5|6}$.} Let us fix $\CN =3$ and consider the direct product $\CP^{3|3}\times\CP^{3|3}_\ast$ of supertwistor space and dual
supertwistor space. The direct product space can be covered by four patches
\begin{equation}\label{3.1}
\hat{\CU}_1:=\hat{\CU}_+\times \hat{\CV}_+~,~~~
\hat{\CU}_2:=\hat{\CU}_-\times \hat{\CV}_-~,~~~
\hat{\CU}_3:=\hat{\CU}_+\times \hat{\CV}_-~,~~~
\hat{\CU}_4:=\hat{\CU}_-\times \hat{\CV}_+~,
\end{equation}
where $\hat{\CU}_\pm$ and $\hat{\CV}_\pm$ are coordinate patches on $\CP^{3|3}$ and $\CP^{3|3}_*$, respectively.  Coordinates on  $\hat{\CU}_a$, 
$a=1,...,4$, are given by those on\   $\hat{\CU}_\pm\subset\CP^{3|3}$ and on\ $\hat{\CV}_\pm\subset\CP_*^{3|3}$. We denote them by
$(z^\alpha_{(a)}, w^\ald_{(a)}, \lambda_\ald^{(a)}$, $\mu_\alpha^{(a)},$ $\eta_i^{(a)},$ $\theta^i_{(a)})$. On nonempty intersections of patches 
\eqref{3.1} these coordinates are transformed in an obvious way which follows from the transformations of coordinates on $\CP^{3|3}$ and $\CP^{3|3}_*$.

Let us consider a {\it quadric} $\CL_{\FC}^{5|6}$ in $\CP^{3|3}\times\CP^{3|3}_*$ defined by the equation~\cite{19}
\begin{equation}\label{3.2}
\omega^\alpha\mu_\alpha-\sigma^\ald\lambda_\ald+2\theta^i\eta_i=0
\end{equation} 
in homogeneous coordinates $(\omega^\alpha , \lambda_\ald , \eta_i)$ and $(\sigma^\ald , \mu_\alpha , \theta^i)$ on $\CP^{3|3}$ and $\CP^{3|3}_*$, 
respectively, and by the equations
\begin{equation}\label{3.3}
z^\alpha_{(a)}\mu_\alpha^{(a)}-w_{(a)}^\ald\lambda_\ald^{(a)}+
2\theta_{(a)}^i\eta_i^{(a)}=0
\end{equation}
on four patches $\hat{\CU}_a$ defined in \eqref{3.1}. The quadric surface  \eqref{3.2}, \eqref{3.3} in $\CP^{3|3}\times\CP^{3|3}_*$ is called the {\it 
superambitwistor space}. It is covered by four patches
\begin{equation}\label{3.4}
\CU_a:=\hat{\CU}_a\cap\CL^{5|6}_\FC~,
\end{equation}
so that $\CL^{5|6}_\FC =\bigcup_{a=1}^4\,\CU_a$.

\medskip

\noindent
{\bf   Fibration of $\CL_{\FC}^{5|6}$ over $\CPP^1\times\CPP^1_*$.} Note that one can introduce a holomorphic projection
\begin{equation}\label{3.5}
q:~\CL^{5|6}_\FC\rightarrow\CPP^1\times\CPP^1_*
\end{equation}
with $(3|6)$-dimensional fibres. Then $\CL_{\FC}^{5|6}$ can be identified with a quotient bundle defined in terms of a short exact sequence of vector bundles
\begin{equation}\label{3.6}
0\to\CL_{\FC}^{5|6}\ \to\ \mbox{pr}^*_1\CP^{3|3}\oplus\mbox{pr}^*_2\CP^{3|3}_*\ \stackrel{\nu}{\to}\ \CO(1,1)\ \to\  0\ ,
\end{equation}
where the mapping $\nu$ is defined by \eqref{3.2}. Here 
\begin{equation}\label{3.7}
\mbox{pr}^*_1\CP^{3|3}=\CO(1,0){\otimes}\FC^2\oplus\CO(1,0){\otimes}\Pi\FC^3\ ,
\end{equation}
\begin{equation}\label{3.8}
\mbox{pr}^*_2\CP^{3|3}_*=\CO(0,1){\otimes}\FC^2\oplus\CO(0,1){\otimes}\Pi\FC^3
\end{equation}
and $\CO(m,n)$ are line bundles over $\CPP^1\times\CPP^1_*$ defined as follows. Let $\CO(m)$ denote the holomorphic line bundle $\CO(m)\to\CPP^1$ parametrized
by the first Chern number $m\in\RZ$. Let pr$_1$ and pr$_2$ are the two projections from $\CPP^1\times\CPP^1_*$ onto $\CPP^1$ and $\CPP^1_*$. Then one introduces
the line bundle over $\CPP^1\times\CPP^1_*$ as
\begin{equation}\label{3.9}
\CO(m,n)=\mbox{pr}^*_1\CO(m)\otimes\mbox{pr}^*_2\CO(n)
\end{equation}
and these bundles are used in \eqref{3.6}-\eqref{3.8}.

\medskip

\noindent
{\bf   Coordinates on $\CPP^1\times\CPP^1_*$.} The base $\CPP^1\times\CPP^1_*$ of the fibration \eqref{3.5} is covered by four patches
\begin{equation}\label{3.10}
V_a\,:=\,\CU_a\cap(\CPP^1\times\CPP^1_*)
\end{equation}
with coordinates 
\begin{equation}\label{3.11}
(\lambda_{(a)},\mu_{(a)}):\quad (\lambda_+,\mu_+)\ , \ \ (\lambda_-,\mu_-)\ , \ \ (\lambda_+,\mu_-)\ \ \mbox{and}  \ \ (\lambda_-,\mu_+)
\end{equation}
 on patches
\begin{equation}\label{3.12}
V_1 = U_+\times V_+\ , \ \  V_2 = U_-\times V_-\ , \ \ V_3 = U_+\times V_- \ \ \mbox{and}  \ \  V_4 = U_-\times V_+\ .
\end{equation}
We denote by $z^A_{(a)}$ with $A=1,2,3$ bosonic coordinates on the fibres over $V_a$ in the bundle \eqref{3.5}. Their explicit expression  in terms of 
$z^\alpha_{(a)}$ and $w^\ald_{(a)}$ can be found e.g. in \cite{16}. Additionally, we use odd variables $\theta^i_{(a)}$ and $\eta_i^{(a)}$ as the 
fermionic coordinates on fibres of the projection $q$ in \eqref{3.5}.

\medskip

\noindent
{\bf  Moduli of complex submanifolds.} Holomorphic sections over $V_a$ of the bundle \eqref{3.5} are spaces $Q^2_{\hat x}\cong\CPP^1\times\CPP^1_*$, 
$\hat x = (x, \theta , \eta )\in \FC^{4|12}$, defined by the equations \cite{19}
\begin{equation}\nonumber
z^\alpha_{(a)}=(x^{\alpha\ald}-\theta^{\alpha i}\eta_i^\ald) \lambda_\ald^{(a)}~,~~~
w^\ald_{(a)}=(x^{\alpha\ald}+\theta^{\alpha i}\eta_i^\ald)\mu_\alpha^{(a)}~,
\end{equation}
\begin{equation}\label{3.13}
\theta^i_{(a)}=\theta^{\alpha i}\mu_\alpha^{(a)}~,~~~
\eta^{(a)}_i=\eta_i^\ald\lambda_\ald^{(a)}~~~\mbox{with}
~~~(\lambda_\ald^{(a)},\mu_\alpha^{(a)})\in V_a~,
\end{equation}
where
\begin{equation}\nonumber
\lambda_\ald^{(1)}=\lambda_\ald^{+}\ ,\ \  \mu_\alpha^{(1)}=\mu_\alpha^{+}\ ,\ \  \lambda_\ald^{(2)}=\lambda_\ald^{-}\ ,\ \  
\mu_\alpha^{(2)}=\mu_\alpha^{-}\ , 
\end{equation}
\begin{equation}\label{3.14}
\lambda_\ald^{(3)}=\lambda_\ald^{+}\ ,\ \  \mu_\alpha^{(3)}=\mu_\alpha^{-}\ ,\ \ 
\lambda_\ald^{(4)}=\lambda_\ald^{-}\ ,\ \  \mu_\alpha^{(4)}=\mu_\alpha^{+}\ .
\end{equation}
By definition, \eqref{3.13} satisfy equations \eqref{3.3} and we may choose three independent functions from $z^\alpha_{(a)}, w^\ald_{(a)}$ and raise them to the 
coordinates $z^A_{(a)}$ \cite{16}. The local sections \eqref{3.13} are properly glued on $V_a\cap V_b\ne\varnothing$ and define a global holomorphic section of the bundle
\eqref{3.5} parametrized by supermoduli $(x, \theta , \eta )\in \FC^{4|12}$.

Equations \eqref{3.13} define a correspondence between $\CL_{\FC}^{5|6}$ and $\FC^{4|12}$ via a double fibration 
\begin{equation}\label{3.15}
\begin{picture}(80,40)
\put(0.0,0.0){\makebox(0,0)[c]{$\CL^{5|6}_\FC$}}
\put(64.0,0.0){\makebox(0,0)[c]{$\FC^{4|12}$}}
\put(32.0,33.0){\makebox(0,0)[c]{$\CF^{6|12}_\FC$}}
\put(7.0,18.0){\makebox(0,0)[c]{$\pi_2$}}
\put(55.0,18.0){\makebox(0,0)[c]{$\pi_1$}}
\put(25.0,25.0){\vector(-1,-1){18}}
\put(37.0,25.0){\vector(1,-1){18}}
\end{picture}
\end{equation}
where $\CF^{6|12}_\FC:=\FC^{4|12}\times\CPP^1\times\CPP^1_*$. Namely, a
point $\hat x=(x,\theta,\eta)\in\FC^{4|12}$ corresponds to
\begin{equation}\label{3.16}
Q^2_{\hat x}=\pi_2(\pi_1^{-1}(\hat x))\cong\CPP^1\times\CPP^1_*
\hra\CL^{5|6}_\FC
\end{equation}
and a point ${\mathfrak l}\in\CL^{5|6}_\FC$ corresponds to a $(1|6)$-dimensional {\it super null line} $\pi_1(\pi_2^{-1}({\mathfrak l}))$ in $\FC^{4|12}$
\begin{equation}\label{3.17}
x^{\alpha\ald}=x_0^{\alpha\ald}+t\mu^\alpha\lambda^\ald\ ,\ \ \eta^\ald_i=\eta^\ald_{0i}+t\beta_i\lambda^\ald\ \ \mbox{and}\ \ 
\theta^{\alpha i}=\theta^{\alpha i}_0+t \gamma^i\mu^\alpha\ ,\ t\in\FR\ ,
\end{equation}
where parameters $\beta_i$ and $\gamma^i$ are odd. The supermanifold $\CF_\FC^{6|12}$ is covered by four patches
\begin{equation}\label{3.18}
\tilde{\CU}_a\,=\,\FC^{4|12}\times V_a
\end{equation}
with coordinates
$(x^{\alpha\ald}, \theta^{\alpha i}, \eta^\ald_i, \lambda_\ald^{(a)}, \mu_\alpha^{(a)})$ on $\tilde{\CU}_a$.

\medskip

\noindent
{\bf  Integrable distribution $\CT$ on $\CF_\FC^{6|12}$.} The tangent spaces to the complex $(1|6)$-dimensional leaves of the fibration
\begin{equation}\label{3.19}
\pi_2:~\CF^{6|12}_\FC\rightarrow\CL^{5|6}_\FC
\end{equation}
from the diagram \eqref{3.15} are spanned by the holomorphic vector fields \cite{19}
\begin{equation}\label{3.20}
W_{(a)}:=\mu_{(a)}^\alpha\lambda_{(a)}^\ald\dpar_{\alpha\ald}~,\quad \dpar_{\alpha\ald}:=\frac{\dpar}{\dpar x^{\alpha\ald}}\ ,
\end{equation}
\begin{equation}\label{3.21}
D^i_{(a)}=\lambda_{(a)}^\ald D_\ald^i~~~\mbox{and}~~~
D^{(a)}_i=\mu_{(a)}^\alpha D_{\alpha i}~,
\end{equation}
which are properly glued on $\tilde{\CU}_a\cap\tilde{\CU}_b\neq\varnothing$ into global vector fields on $\CF^{6|12}_\FC$. 
Here $D_\ald^i$ and  $D_{\alpha i}$ are odd vector fields 
\begin{equation}\label{3.22}
D^i_{\ald}=\frac{\dpar}{\dpar \eta^{\ald}_i} + \theta^{\alpha i}\dpar_{\alpha\ald}\quad\mbox{and}\quad
D_{\alpha i}=\frac{\dpar}{\dpar\theta^{\alpha i}} + \eta^{\ald}_i\dpar_{\alpha\ald}
\end{equation}
used in the formulation of $\CN=3$ SYM theory (see e.g.~\cite{15,18}). Note that vector fields \eqref{3.20}, \eqref{3.21} define the integrable distribution $\CT$
in the holomorphic tangent bundle of $\CF^{6|12}_\FC$.

\section{From holomorphic bundles to $\CN =3$ super-Yang-Mills}

In this section we describe the twistor procedure for generating solutions of $\CN =3$ SYM field equations from \v{C}ech data defining holomorphic 
bundles $\CE$ over the superambitwistor space $\CL^{5|6}_\FC$. 

\medskip

\noindent
{\bf Bundles $\CE$ over $\CL^{5|6}_\FC$.} For defining a holomorphic rank $N$ vector bundle $\CE$ over $\CL^{5|6}_\FC$, one should fix an open covering
$\{\CU_a\}$ of  $\CL^{5|6}_\FC$, e.g. choose the one that was specified in \eqref{3.10}-\eqref{3.12}, and consider a collection $\{f_{ab}\}$ of {\it holomorphic}
GL$(N, \FC$)-valued functions (\v{C}ech 1-cocycles) defined on $\CU_a\cap\CU_b$ such that
\begin{equation}\label{4.1}
f_{ab}f_{bc}f_{ca}=\unit_N\quad\mbox{on}\quad\CU_a\cap\CU_b\cap\CU_c\neq\varnothing\ .
\end{equation}
We restrict ourselves to topologically trivial bundles $\CE$, i.e. those for which there exists a collection $\{{\psi}_a\}$ of
continuous GL$(N, \FC$)-valued functions (\v{C}ech 0-cochain) defined on $\CU_a$ such that
\begin{equation}\label{4.2}
f_{ab}={\psi}_a^{-1}{\psi}_b
\end{equation}
on any nonempty intersections $\CU_a\cap\CU_b$ of charts. Note that ${\psi}_a$'s are not holomorphic but only $C^0$-functions and therefore $\CE$ is 
trivial only topologically but not holomorphically. However, in the twistor description of gauge theories it is assumed that restriction of the bundle $\CE$ to
any quadric $Q^2_{\hat x}\cong \CPP^1\times\CPP^1_*\hra\CL^{5|6}_\FC$ is holomorphically trivial. This means that $C^0$-functions\footnote{One can also 
consider $C^k$, $C^\infty$ and real analytic GL$(N, \FC )$-valued functions.} $\psi_a$ in \eqref{4.2} holomorphically depend on coordinates 
$\lambda_\ald$ and $\mu_\alpha$ on
$\CPP^1\times\CPP^1_*$. Also, they holomorphically depend on $\eta_i$ and $\theta^i$. Thus, on the superambitwistor space $\CL^{5|6}_\FC$ we have a collection 
$\{f_{ab}\}$ of GL$(N,\FC)$-valued holomorphic functions defined on $\CU_a\cap\CU_b$ and constrained by equations \eqref{4.1}. For them we have a 
generalized Riemann-Hilbert problem \eqref{4.2} of finding GL$(N, \FC )$-valued functions $\psi_a$ holomorphic in coordinates on $Q^2_\hx\hra \CL^{5|6}_\FC$
and continuous or smooth in other coordinates.

\medskip

\noindent
{\bf Bundles $\tilde\CE$ over $\CF^{6|12}_\FC$.} The first step in the twistor approach is to take transition functions $\{f_{ab}\}$ (\v{C}ech data) of holomorphic
bundle $\CE\to\CL^{5|6}_\FC$, resolve the nonlinear 1-cocycle constraints \eqref{4.1} and then find solutions $\{\psi_{a}\}$ of the generalized Riemann-Hilbert 
problem \eqref{4.2}.

The second step is to pull back $\CE$ to the bundle $\tilde\CE =\pi_2^*\CE$ over the supermanifold $\CF^{6|12}_\FC$ with a covering $\{\tilde\CU_{a}\}$ given
by \eqref{3.18}. The pulled back transition functions\footnote{For simplicity, we denote the pulled-back transition functions also by  $f_{ab}$, slightly abusing notation. 
The same holds true for functions $\psi_a$.} $\{f_{ab}\}$ have to be constant along fibres of the bundle \eqref{3.19}, i.e. 
 \begin{equation}\label{4.3}
W_{(a)}f_{ab}=D^i_{(a)}f_{ab}=D_i^{(a)}f_{ab}=0~,
\end{equation}
where vector fields $W_{(a)}$, $D^i_{(a)}$ and $D_i^{(a)}$ are given in \eqref{3.20}-\eqref{3.22}. Recall that these vector fields generate an integrable 
subbundle $\CT$ of the tangent bundle $T\CF^{6|12}_\FC$. Let $\dd_{\CT}$ denote the restriction of the exterior derivative $\dd$ on $\CF^{6|12}_\FC$ to
$\CT$. Then equations \eqref{4.3} can be rewritten as
 \begin{equation}\label{4.4}
\dd_{\CT} f_{ab} =0\ ,
\end{equation}
where $\dd_{\CT}^2=0$ due to integrability of $\CT$. 

So, on the second step we consider functions $f_{ab}$ depending on the coordinates \eqref{3.13} which are the pull-back to  $\CF^{6|12}_\FC$
of coordinate functions on $\CL^{5|6}_\FC$ and impose on  $f_{ab}$  the differential constraint equations \eqref{4.4}. However, we do not impose 
these equations on $\psi_a$ and therefore after substitution of \eqref{4.2} into \eqref{4.4} we obtain 
\begin{align}\label{4.5}
&{\psi}_aD^i_{(a)}{\psi}_a^{-1}=\psi_b D^i_{(a)}\psi^{-1}_b=: A^i_{(a)}=
\lambda^\ald_{(a)}A_\ald^i(x,\theta,\eta)~,\\
\label{4.6}
&{\psi}_aD_i^{(a)}{\psi}_a^{-1}=\psi_b D_i^{(a)}\psi^{-1}_b=:A_i^{(a)}=
\mu^\alpha_{(a)}A_{\alpha i}(x,\theta,\eta)~,\\
\label{4.7}
&{\psi}_a W_{(a)}{\psi}_a^{-1}=\psi_bW_{(a)}\psi^{-1}_b=: A_{w_{(a)}}=\mu^\alpha_{(a)}
\lambda^\ald_{(a)}A_{\alpha\ald}(x,\theta,\eta)~,
\end{align}
where $(x,\theta,\eta)=(x^{\alpha\ald}\cb\theta^{\alpha i}\cb\eta_i^\ald)$. Note that the last equalities in \eqref{4.5}-\eqref{4.7} follow from a
generalized Liouville theorem on $\CPP^1\times \CPP^1_*$ which says that $A_{(a)}^i$ is a section of the bundle $\CO(1,0)$,
$A_i^{(a)}$ is a section of the bundle $\CO(0,1)$ and $A_{w_{(a)}}$ is a section of the bundle $\CO(1,1)$ over
$\CPP^1\times \CPP^1_*$. Equations \eqref{4.5}-\eqref{4.7} are the output of the second step.

\medskip

\noindent
{\bf Bundles $E$ over $\FC^{4|12}$.} The third (and last) step is to push down the bundle $\tilde\CE\to\CF^{6|12}_\FC$ to $\FC^{4|12}$. This means that we
define the vector bundle $E\to\FC^{4|12}$ to be the bundle whose fibre at $\hat x\in \FC^{4|12}$ is the space of global sections of $\tilde\CE|_{Q^2_\hx}
\to Q^2_\hx $ and, as $\hx$ varies, this will form a vector bundle $E=\pi_{1\ast}\pi^\ast_2\CE$ over $\FC^{4|12}$. Note that using the splitting
 \eqref{4.2} of transition functions $f_{ab}$ of the bundle $\tilde\CE$ in \eqref{4.4} means going over to transition functions $\tilde f_{ab}=\unit_N$ and the
 covariant derivative $\nabla_\CT$ (or connection $A_\CT$) on $\tilde\CE$ along the distribution $\CT$ given by \eqref{3.20}-\eqref{3.22}, i.e. 
 $(f_{ab}, \dd_\CT)\to (\tilde f_{ab}=\unit_N , \nabla_\CT)$.
 
 The equations \eqref{4.5}-\eqref{4.7} can be written as the linear system~\cite{12,19}
\begin{equation}\label{4.8}
\nabla_{\CT} \psi_{a} =0\ ,
\end{equation}
which in components read
\begin{align}\label{4.9}
&\mu_{(a)}^\alpha\lambda_{(a)}^\ald(\dpar_{\alpha\ald}+A_{\alpha\ald})\psi_a=:  \mu_{(a)}^\alpha\lambda_{(a)}^\ald \nabla_{\alpha\ald}\psi_a= 0~,\\
\label{4.10}
&\lambda_{(a)}^\ald(D^i_{\ald}+A^i_{\ald})\psi_a=: \lambda_{(a)}^\ald\nabla^i_{\ald}\psi_a=0~,\\
\label{4.11}
&\mu_{(a)}^\alpha(D_{\alpha i}+A_{\alpha i})\psi_a=:\mu_{(a)}^\alpha\nabla_{\alpha i}\psi_a=0~.
\end{align}
Here $\nabla_{\alpha\ald}, \nabla^i_{\ald}$ and $\nabla_{\alpha i}$ are gauge covariant derivatives in the trivial bundle $E=\FC^{4|12}\times\FC^N$
over $\FC^{4|12}$ and $A_{\alpha\ald}$, $A^i_{\ald}$ and $A_{\alpha i}$ are components of the superconnection. Note also that
\begin{equation}\label{4.12}
\nabla_{\CT}=\{\mu_{(a)}^\alpha\lambda_{(a)}^\ald \nabla_{\alpha\ald}\, ,\, \lambda_{(a)}^\ald\nabla^i_{\ald}\, ,\, \mu_{(a)}^\alpha\nabla_{\alpha i}\}
\end{equation}
is the covariant derivative along $(1|6)$-dimensional null lines in $\FC^{4|12}$. 

The linear system \eqref{4.9}-\eqref{4.11} has been known for a long time~\cite{13, 18}. Its compatibility conditions are the constraint equations
\begin{equation}\label{4.13}
\nabla^2_\CT=0\ \Leftrightarrow\ \{\nabla_\ald^i,\nabla_\bed^j\}{+}\{\nabla_\bed^i,\nabla_\ald^j\}{=}0,~
\{\nabla_{\alpha i},\nabla_{\beta j}\}{+}\{\nabla_{\beta i},\nabla_{\alpha j}\}{=}0,~
\{\nabla_{\alpha i},\nabla^j_{\ald}\}{-}2\delta_i^j\nabla_{\alpha\ald}{=}0,
\end{equation}
 of $\CN{=}3$ SYM theory. This means vanishing of the supercurvature along $(1|6)$-dimensional null lines in $\FC^{4|12}$. Recall that these
 equations are equivalent to the field equations of $\CN{=}3$ SYM theory~\cite{13,15,18}. Using the expansions of the superconnection in the odd 
 variables, one can rewrite \eqref{4.13} as equations on a supermultiplet of ordinary fields. Moreover, these equations turn out to be equivalent
 to the equations of motion of  $\CN{=}4$ SYM theory in complexified Minkowski space $\FC^{4}$. In fact, $\CN{=}3$ and $\CN{=}4$ SYM theories have the same physical content, i.e. they are on-shell equal.
 
 Note that the twistor correspondence can be considered for any number $\CN{\le}3$ supersymmetries~\cite{15,18}.  To do this, one should replace the three spaces in the double fibration  \eqref{3.15} with the spaces $\CL_{\FC}^{5|2\CN}$, $\CF_{\FC}^{6|4\CN}$ and $\FC^{4|4\CN}$. After this replacement, everything remains valid, and we obtain a correspondence between holomorphic vector bundles $\CE$ over $\CL_{\FC}^{5|2\CN}$ and solutions of the constraint equations  \eqref{4.13} with $i,j=1,...,\CN\le 3$. These equations mean that the gauge supercurvature on $\FC^{4|4\CN}$ vanishes along $(1|2\CN )$-dimensional  null lines in $\FC^{4|4\CN}$. For $\CN<3$ equations \eqref{4.13} are not equivalent to $\CN$-extended SYM equations.  In the bosonic $\CN =0$ case,  the constraint equations disappear.  
 
\medskip\noindent{\bf A special gauge.} Note that restriction of a vector bundle $\CE\rightarrow\CL_{\FC}^{5|6}$ to the fibres
$\FC^{3|6}_{\lambda,\mu}$ of the bundle \eqref{3.5} are holomorphically trivial since all these fibres are contractible. Therefore, there exist trivializations
$\tilde{\psi}_a$ of $\CE$ over $\CU_a$ such that
\begin{equation}\label{4.14}
f_{ab}={\psi}^{-1}_a{\psi}_b=\tilde{\psi}^{-1}_a\tilde{\psi}_b~~~
\mbox{on}~~~\CU_a\cap\CU_b\neq\varnothing\ .
\end{equation}
It can be shown \cite{19} that the linear system \eqref{4.8} can be  transformed to
the system
\begin{align}\label{4.15}
\dpar_{\bar{z}_{(a)}^A}\tilde{\psi}_a&=0~,\\\label{4.16}
(\dpar_{\bl_{(a)}}+\tilde{\CA}_{\bl_{(a)}})\tilde{\psi}_a&=0~,\\
\label{4.17}
(\dpar_{\bar\mu_{(a)}}+\tilde{\CA}_{\bar\mu_{(a)}})\tilde{\psi}_a&=0~,
\end{align}
where $z^A_{(a)}$ with $A=1,2,3$ are bosonic coordinates on fibres $\FC^{3|6}_{\lambda,\mu}$ and $\tilde{\CA}_{\bl_{(a)}}$, $\tilde{\CA}_{\bar\mu_{(a)}}$ are related with components in \eqref{4.5}-\eqref{4.7} by a gauge transformation. In this gauge the compatibility conditions are the equations
\begin{equation}
\dpar_{\bar{z}_{(a)}^A}\tilde{\CA}_{\bl_{(a)}}=0=\dpar_{\bar{z}_{(a)}^A}\tilde{\CA}_{\bar\mu_{(a)}}\quad\mbox{and}\quad
\dpar_{\bl_{(a)}}\tilde{\CA}_{\bar\mu_{(a)}}-\dpar_{\bar\mu_{(a)}}\tilde{\CA}_{\bl_{(a)}}+[\tilde{\CA}_{\bl_{(a)}},\tilde{\CA}_{\bar\mu_{(a)}}]=0
\end{equation}
which are equivalent to the constraint equations  \eqref{4.13}. These equations were discussed in \cite{DOg} and many other papers.

\medskip
\noindent
{\bf Non-integrability of SYM on $\FC^{4}$.}  Summarizing, we have described the twistor procedure for solving $\CN{=}3$ SYM field equations. 
It consists of the following steps:
\begin{itemize} 
\item One takes GL$(N, \FC)$-valued holomorphic functions $f_{ab}$ on subsets $\CU_a\cap\CU_b$ of the superambitwistor space $\CL^{5|6}_\FC$.
These functions have to satisfy the nonlinear algebraic constraints \eqref{4.1} (difficult to resolve) and the equations \eqref{4.4} of constancy along the
distribution $\CT$ (easy to resolve).
\item One should solve a generalized Riemann-Hilbert problem  \eqref{4.2} on $Q^2_{\hx}\cong\CPP^1{\times}\CPP^1_*$ (the space of 
spectral parameters) for $\hx$ in an open subset of $\FC^{4|12}$.
\item One defines a superconnection $A_\CT$ by using $\psi_a$ and formulae \eqref{4.5}-\eqref{4.7}. By twistor construction, $A_\CT$ satisfies 
the zero-curvature equations \eqref{4.13}, which are equivalent to field equations of $\CN{=}3$ SYM theory.
 \end{itemize} 
Thus, there are a linear system \eqref{4.8}  and a zero curvature representation \eqref{4.13} for the field equations  of $\CN{=}3$ SYM theory on complexified 
Minkowski space. However, one cannot call this model integrable since solutions of SYM equations \eqref{4.13} are encoded into the twistor data
$\{f_{ab}\}$ which are {\it not free} due to 1-cocycle constraints \eqref{4.1}. It is instructive to compare this with the self-dual subsector of 
$\CN{=}3$ SYM theory, which is integrable because the space of spectral parameters there is the Riemann sphere $\CPP^1$ covered by two
open sets $U_\pm$ and hence the twistor construction is reduced to the Riemann-Hilbert problem \eqref{2.18} for {\it one}  matrix-valued function 
$f_{+-}(\hx , \lambda)$
(free data; no constraints \eqref{4.1}). In the next sections we will show that $\CN{=}3$ SYM model becomes integrable after imposing reality
conditions reducing the model to Minkowski space $\FR^{3,1}$. For this, it is necessary to analyze complexifications of real analytic manifolds.

\section{Minimal and non-minimal complexifications of $S^2$}

\noindent
{\bf Quadric $Q^2_\FC$.} We want to describe an ambitwistor correspondence for $\CN{=}3$ SYM theory on the Minkowski space $\FR^{3,1}$. To do this,
one should understand how null lines in the real case are related to the null lines in the complex case.

Recall that the vector field
\begin{equation}\label{5.1}
W=\mu^\alpha\lambda^\ald\dpar_{\alpha\ald}=W^\mu\dpar_\mu\ ,
\end{equation}
introduced in \eqref{3.20}, defines tangent vector to null lines in $\FC^4$ for any value of $\mu^\alpha$ and $\lambda^\ald$ in $\FC^2\backslash\{0\}$. Here 
$[\lambda_\ald]$ and $[\mu_\alpha]$ are projective coordinates on $\CPP^1$ and $\CPP^1_*$, respectively. Components $W^\mu$ of the null vector field
\eqref{5.1} are 
\smallskip
\begin{equation}\label{5.2}
\left.\begin{array}{lr}
W^0=\mu^1\lambda^{\dot 1}+\mu^2\lambda^{\dot 2}=\mu_1\lambda_{\dot 1}+\mu_2\lambda_{\dot 2}\ ,&
W^1=\mu^1\lambda^{\dot 2}+\mu^2\lambda^{\dot 1}=-(\mu_1\lambda_{\dot 2}+\mu_2\lambda_{\dot 1})\ ,
\\
W^2=\di(\mu^1\lambda^{\dot 2}-\mu^2\lambda^{\dot 1})=\di(\mu_1\lambda_{\dot 2}-\mu_2\lambda_{\dot 1})\ ,&
W^3=\mu^1\lambda^{\dot 1}-\mu^2\lambda^{\dot 2}=-(\mu_1\lambda_{\dot 1}-\mu_2\lambda_{\dot 2})\ .
\end{array}\right.
\end{equation}
It is easy to check that
\begin{equation}\label{5.3}
\eta_{\mu\nu} W^\mu W^\nu = -(W^0)^2 +  (W^1)^2 +(W^2)^2 +(W^3)^2 =0
\end{equation} 
and the embedding $\CPP^1{\times}\CPP^1_*\to\CPP^3$ defined by \eqref{5.2} induces an isomorphism
\begin{equation}\label{5.4}
\CPP^1{\times}\CPP^1_*\cong Q^2_\FC\ ,
\end{equation} 
where $ Q^2_\FC$ is the quadric surface in $\CPP^3$ defined by the equation \eqref{5.3}. Here $\{W^\mu\}$ are projective coordinates on $\CPP^3$.

\medskip

\noindent
{\bf Quadric $Q^2_\FR$.} Recall that a real structure on a complex manifold $X_\FC$ is defined as an antiholomorphic involution $\rho :\ X_\FC\to X_\FC$.
For the Minkowski signature the anti-linear transformations of dotted and undotted spinors read (see e.g. \cite{12,19})
\begin{equation}\label{5.5}
\rho (\lambda_\ald )=\overline{\mu_\alpha}\quad\mbox{and}\quad\rho (\mu_\alpha)=\overline{\lambda_\ald }\ .
\end{equation} 
Here, $[\lambda_\ald ]$ and $[\mu_\alpha ]$ are homogeneous coordinates on two Riemann spheres and the involution $\rho$ maps these spheres one
into another. 

The fixed point set of the map $\rho : \CPP^1{\times}\CPP^1_*\to\CPP^1{\times}\CPP^1_*$ forms the two-sphere \cite{28}
\begin{equation}\label{5.6}
S^2=\CPP^1=\diag (\CPP^1\times\overline{\CPP^1})\ ,
\end{equation} 
where $\overline{\CPP^1} (=\CPP^1_*)$ denotes the Riemann sphere with the opposite complex structure. On the two-sphere \eqref{5.6} we have 
$\mu_\alpha=\overline{\lambda_\ald }$ and therefore all four coordinates $W^{\mu}$ of a null vector in the real Minkowski space $\FR^{3,1}$ will be
real,
\begin{equation}\label{5.7}
W^0=\lambda_{\dot 1}\overline{\lambda_{\dot 1}}+\lambda_{\dot 2}\overline{\lambda_{\dot 2}},\ 
W^1=-(\lambda_{\dot 1}\overline{\lambda_{\dot 2}}+\lambda_{\dot 2}\overline{\lambda_{\dot 1}}),\ 
W^2=\di (\lambda_{\dot 2}\overline{\lambda_{\dot 1}}-\lambda_{\dot 1}\overline{\lambda_{\dot 2}}),\ 
W^3=\lambda_{\dot 2}\overline{\lambda_{\dot 2}}-\lambda_{\dot 1}\overline{\lambda_{\dot 1}}\ .
\end{equation} 
Hence, the equation \eqref{5.3} for the real projective coordinates $\{W^\mu\}$ on $\FR P^{3}$ will define a real quadric $Q^2_\FR$ in $\FR P^{3}$.

Note that the subspace $\Gamma=\{W^0=0\}$  in $\FR P^{3}$ does not intersect the quadric \eqref{5.3} since $\lambda_\ald \in \FC^2{\setminus}\{0\}$ and 
therefore we can introduce real coordinates 
\begin{equation}\label{5.8}
x^i = \frac{W^i}{W^0}\quad\mbox{for}\quad i=1,2,3\ ,
\end{equation} 
obeying the equation
\begin{equation}\label{5.9}
\left\|x\right\|^2:=\delta_{ij}x^ix^j=1\ ,
\end{equation} 
which follows from \eqref{5.3}. Thus, the inclusion $S^2\to Q^2_\FR$ is a diffeomorphism
\begin{equation}\label{5.10}
S^2\cong Q^2_\FR\ ,
\end{equation} 
i.e. $S^2$ can be considered as a quadric in $\FR P^{3}$.

\medskip

\noindent
{\bf Complex sphere $S^2_\FC$.}  Note that the notion of complexification of manifolds is not uniquely defined (see e.g.~\cite{29,30} and references therein). On a real
analytic manifold $X$ one can always define a complex structure on the tangent bundle $TX\cong X_\FC$ (on a neighbourhood of zero section) and this is a {\it minimal 
complexification} \cite{29,30}. For the minimal complexification the inclusion $X\to X_\FC$ is a {\it homotopy equivalence}. This is not so for the inclusion of $Q_\FR^2=S^2$ 
into $Q_\FC^2=\CPP^1{\times}\CPP^1_*$, i.e. $Q_\FC^2$ is not a minimal complexification of $S^2$.

The minimal complexification $S^2_\FC$ of $S^2$ is described as follows. Consider the real two-sphere $S^2$ as the surface \eqref{5.9} embedded into $\FR^3$.
Then the complexification of $S^2$ is defined by the equations
\begin{equation}\label{5.11}
\delta_{ij}z^iz^j=1\quad \Leftrightarrow\quad\left\|x\right\|^2-\left\|y\right\|^2=1\quad\mbox{and}\quad\delta_{ij}x^iy^j=0\quad\mbox{for}\quad z^i=x^i+\di y^i\ .
\end{equation} 
Note that $\Gamma=\{W^0=0\}$ is an irreducible curve in $Q_\FC^2\subset\CPP^3$ and on $Q_\FC^2\setminus\Gamma$ we can divide all
terms of the equation \eqref{5.3} by $(W^0)^2$ obtaining \eqref{5.11} for complex coordinates
\begin{equation}\label{5.12}
z^i:=\frac{W^i}{W^0}\ .
\end{equation} 
Hence, we have an isomorphism
\begin{equation}\label{5.13}
S^2_\FC\cong Q^2_\FC\setminus\Gamma\ ,
\end{equation} 
i.e. the minimal complexification $S^2_\FC$ of $S^2$ is the submanifold of $Q_\FC^2=\CPP^1{\times}\CPP^1_*$ covered by two patches, $U_+\times V_+$ and 
$U_-\times V_-$, of four patches given in \eqref{3.12}.

\noindent
{\bf Tangent bundle $TS^2$.}
Let us now establish a diffeomorphism of  $S^2_\FC$ with the tangent bundle $TS^2$ mentioned at the beginning of the previous subsection. The manifold $TS^2$ is defined by the equations 
\begin{equation}\label{5.14}
\left\|X\right\|^2 = \delta_{ij} X^iX^j=1\quad\mbox{and}\quad\delta_{ij}X^iY^j=0\ .
\end{equation}
It is easy to see that the map, defined by
\begin{equation}\label{5.15}
z:\ TS^2\ \to\ S^2_\FC\ ,\quad z^i(X,Y)=X^i\sqrt{1+\left\|Y\right\|^2}+\di Y^i\ ,
\end{equation}
is a diffeomorphism with the inverse map
\begin{equation}\label{5.16}
h:\ S^2_\FC\ \to\ TS^2\ ,\quad h(z)=\left(\frac{x^i}{\sqrt{1+\left\|y\right\|^2}}, y^i\right)\quad\mbox{for}\quad z^i=x^i + \di y^i\ .
\end{equation}
Hence, $S^2_\FC\cong TS^2$ is homotopy equivalent to $S^2$, covered by two patches $U_+$ and $U_-$ and it is a minimal complexification of $S^2$.
On the other hand, $\CPP^1{\times}\CPP^1_*\cong Q_\FC^2$ is a non-minimal complexification of $\CPP^1$ which can be defined for any {\it complex}
manifold, $X_\FC\to X_\FC\times \bar  X_\FC$ \cite{28}.

\section{Superambitwistors for Minkowski signature}

\noindent
{\bf Reality conditions.} Recall that a real structure on a complex (super)manifold $X_\FC$ is defined as an antiholomorphic involution $\rho : X_{\FC}\to X_\FC$.
Its fixed locus is the space $X$ of real points, $\rho (X)=X$, of $X_\FC$ (which may be empty). For the real Minkowski space $\FR^{3,1}\subset\FC^4$ the 
antiholomorphic involution on $\CP^{3|3}\times\CP^{3|3}_*$ is defined as follows (see e.g. \cite{12, 19}):
\begin{equation}\label{6.1}
\rho (\omega^\alpha,\lambda_\ald,\eta_i;\mu_\alpha,\sigma^\ald,\theta^i)=
(-\overline{\sigma^\ald},\overline{\mu_\alpha},\overline{\theta^i};
\overline{\lambda_\ald},-\overline{\omega^\alpha},\overline{\eta_i})\ .
\end{equation}
For $\FC^{4|12}$ we have
\begin{equation}\label{6.2}
\rho (x^{\alpha\dot\beta})=-\bar x^{\beta\ald}\quad\mbox{and}\quad\rho(\eta^\ald_i)=\overline{\theta^{\alpha i}}\ .
\end{equation}
Hence, on real slices in $\FC^{4|12}$ and $\CF^{6|12}_\FC$ we have 
\begin{equation}\label{6.3}
\bar x^{\beta\ald}=-x^{\alpha\dot\beta}\ ,\quad\theta^{\alpha i}=\overline{\eta^\ald_i}\quad\mbox{and}\quad\mu_\alpha=\overline{\lambda_\ald}\ .
\end{equation}
If we choose
\begin{equation}\label{6.4}
x^{1\dot{1}}=-\di(x^0+x^3),~
x^{1\dot{2}}=-\di(x^1-\di x^2),~
x^{2\dot{1}}=-\di(x^1+\di x^2),~
x^{2\dot{2}}=-\di(x^0-x^3)
\end{equation}
with $x^\mu\in \FR^4$ for $\mu =0,...,3$, then
\begin{equation}\label{6.5}
\dd s^2=\det(\dd x^{\alpha\ald})=\eta_{\mu\nu}\dd x^{\mu}\dd x^{\nu}\ ,\ \ \eta =\diag(-1, 1, 1, 1)
\end{equation}
is a metric on $\FR^{3,1}$. Thus, coordinates $x^\mu$, $\theta^{\alpha i}$, $\overline{\theta^{\alpha i}}$ parametrize the $\CN =3$ Minkowski superspace
 $\FR^{4|12}\subset\FC^{4|12}$ and $(x, \theta , \bar\theta , \lambda , \bar\lambda )$ are coordinates on $\CF^{6|12}_\FR=\FR^{4|12}\times S^2\subset
 \CF^{6|12}_\FC$. 

The fixed point set of involution \eqref{6.1} is the diagonal $\diag(\CP^{3|3}\times\bar\CP^{3|3})$ of real dimension $(6|6)$. This involution also picks a 
real quadric hypersurface $\CL_\FR^{5|6}$ in $\diag(\CP^{3|3}\times\bar\CP^{3|3})$ fibred over $S^2$, 
\begin{equation}\label{6.6}
\CL_\FR^{5|6}\to S^2\ ,
\end{equation}
where $S^2$ is defined in \eqref{5.6}. The quadric $\CL_\FR^{5|6}$ is the subset of fixed points of the involution $\rho : \CL_\FC^{5|6}\to\CL_\FC^{5|6}$
defined by  \eqref{6.1}. Thus, we obtain a real version of the double fibration \eqref{3.15},
\smallskip
\begin{equation}\label{6.7}
\begin{picture}(80,40)
\put(0.0,0.0){\makebox(0,0)[c]{$\CL_{\FR}^{5|6}$}}
\put(64.0,0.0){\makebox(0,0)[c]{$\FR^{4|12}$}}
\put(35.0,37.0){\makebox(0,0)[c]{$\CF_{\FR}^{6|12}$}}
\put(7.0,20.0){\makebox(0,0)[c]{$\pi_2$}}
\put(55.0,20.0){\makebox(0,0)[c]{$\pi_1$}}
\put(25.0,27.0){\vector(-1,-1){18}}
\put(37.0,27.0){\vector(1,-1){18}}
\end{picture}
\end{equation}
\smallskip
The dimensions of all spaces in this diagram are real. Moreover, spaces $S^2$, $\FR^{4|12}$, $\CF^{6|12}_\FR$, $\CL_\FR^{5|6}$ and 
$\diag(\CP^{3|3}\times\bar\CP^{3|3}_*)$ are {\it totally real} Cauchy-Riemann subspaces of complex spaces $\CPP^1\times\CPP^1_*$,
$\FC^{4|12}$, $\CF^{6|12}_\FC$, $\CL_\FC^{5|6}$ and $\CP^{3|3}\times\CP^{3|3}_*$ , respectively. 

\medskip

\noindent
{\bf Remark.} In the purely bosonic case, one can introduce real U($N$)-valued gauge fields on $\FR^4$ with Minkowski signature (3,1), Kleinian signature (2,2) 
or Euclidean signature(4,0) by choosing an appropriate real structure on $\FC^4$ (see e.g. \cite{19} for a discussion). Ambitwistor spaces for all these cases are
different submanifolds of $\CL^5_\FC\cong\CL^{10}_\FR$ - a complex ambitwistor space with real dimension ten. In the Minkowski case the real ambitwistor
space $\CL^5_\FR$ is a fibre bundle over $S^2$, cf.  \eqref{6.6}. In the Kleinian case $\CL^5_\FR\subset\CL^{5}_\FC$ is a real 5-manifold fibred over a torus 
$S^1\times S^1_*$ \cite{19}. In the Euclidean case, the ambitwistor space is a 8-dimensional Cauchy-Riemann (CR) submanifold $\CL^8_\FR$ of 
$\CL^{10}_\FR\cong\CL^5_\FC$ fibred over $\CPP^{1}\times\CPP^{1}_*$ and having CR rank $m=3$ \cite{31}. In the Minkowski and Kleinian cases,
ambitwistor spaces are embedded 5-dimensional CR submanifolds of $\CL^5_\FC$ with zero CR ranks.

\medskip

\noindent
{\bf Cauchy-Riemann structure.} For a moment, let us restrict ourselves to the purely bosonic setup. A {\it Cauchy-Riemann (CR) structure} on a smooth manifold 
$X$ of real dimension $m$ is a complex subbundle $\CD$ of rank $r$ of the complexified tangent bundle $T^\FC X$ such that $\CD\cap\bar\CD =\{0\}$ and $\CD$
is involutive (integrable), i.e. the space of smooth sections of $\CD$ is closed under the Lie bracket for vector fields. Obviously, the distribution $\bar\CD$ is
integrable if $\CD$ is integrable. The pair $(X, \bar\CD )$ is called a CR manifold  of dimension $m=\dim_\FR X$, of rank $r=\dim_\FC\bar\CD$ and of CR 
codimension $m-2r$. In particular, a CR structure on $X$ in the special case $m=2r$ is a complex structure on $X$ and in general $X$ is a {\it partially complex 
manifold}. Thus, the notion of CR manifolds generalizes that of complex manifolds.

\medskip

\noindent
{\bf Embedded CR manifolds.} The central objects in the theory of CR manifolds are real submanifolds of complex manifolds. They are defined as follows \cite{32}.
Let $(M,J)$ be a $2n$-dimensional smooth real manifold with an integrable almost complex structure $J$, i.e. $M_\FC = (M,J)$ is a complex manifold of
dimension $n=\dim_\FC M_\FC$. If a real $m$-dimensional manifold $X$ is embedded in $M_\FC$ then the tangent bundle $TX$ is embedded in $TM_\FC$
and $TX\subset TM_{\FC|X}$. For any point $x\in X$ the tangent space $T_xX$ is a real subspace in the $\FC$-linear space $T_xM_\FC$. It is clear that
$\CD_x:=T_xM\cap J(T_xM)$ for  $M_\FC=(M,J)$ is a maximal $\FC$-linear subspace in $T_xM_\FC$ belonging to $T_xX$. It is called the complex tangent 
space to $X$ at the point $x\in X$. Its complex dimension $r$ is called the CR-dimension (or rank) of $X$ at the point $x$. If $r=\dim_\FC\CD_x$ is constant on $X$
then $\CD:=\{(x, \xi ): x\in X, \xi\in\CD_x\}$ is a subbundle in $TX$ with the complex structure $J_{x|\CD_x}$ in fibres. Such manifolds $X$ are embedded 
CR manifolds with {\it induced CR structure}.

For the embedded CR manifold $X$ we have $m=2r+k$, where $0\le k\le m$ is a CR codimension of $X$. For $k=0 \Leftrightarrow m=2r$ the manifold $X$
is a complex submanifold of $M_\FC$. In another special case, $r=0\Leftrightarrow m=k$, $X$ is called a {\it totally real submanifold} of $M_\FC$. Of course,
$m=2n-p$ with $p\ge 0$ for $X$ embedded in $M_\FC$ with $\dim_\FC M_\FC =n$. Examples of totally real submanifolds are $\FR^n$ in $\FC^n$, torus 
$|z_1|=...=|z_n|=1$ in $\FC^n$ and subspaces $X$ in $M_\FC$ ($S^2$, $\FR^{4|12}$ etc.) listed after the diagram \eqref{6.7}. Note also that any complex 
manifold $X_\FC$ can be embedded as a totally real submanifold into $M_\FC=X_\FC\times\bar X_\FC$ \cite{28}. We saw this on the examples of 
$\CPP^1=\diag (\CPP^1\times
\overline{\CPP}^1)$ and $\CP^{3|3}=\diag(\CP^{3|3}\times\bar\CP^{3|3})$. The complex structure on $X_\FC$ for such embeddings is not induced from 
the complex structure on $X_\FC\times\bar X_\FC$ and therefore holomorphic bundles over $X_\FC\times\bar X_\FC$ are reduced to real analytic bundles
over the totally real submanifold $X_\FC\subset X_\FC\times\bar X_\FC$. In a twistor context, CR manifolds were considered e.g. in 
description of supersymmetric monopoles~\cite{33} and in an ambitwistor description of $\CN=3$ SYM theory on Euclidean space $\FR^{4,0}$ \cite{31}. Holomorphic
vector bundles of twistor approach become CR vector bundles over CR supermanifolds and they are real analytic complex vector bundles over totally real
CR manifolds.

\medskip

\noindent
{\bf Covering of $\CL_\FR^{5|6}$.} For any complex $n$-dimensional manifold $M_\FC$ it is possible to choose an open covering $\CU=\{\CU_a\}_{a\in I}$ 
such that each nonempty finite intersection of open sets $\CU_a$ is diffeomorphic to a Stein manifold. We described such coverings for $\CF^{6|12}_\FC$, 
$\CL^{5|6}_\FC$ and $\CPP^{1}\times\CPP^{1}_*$ in section 3. They consist of four patches, and the coverings of real subspaces in the above spaces consist
of those patches that belong to fixed point sets of real structures  $\rho$ defined on these spaces. In sections 3 and 5 we saw that $S^2=\diag(\CPP^1\times
\overline{\CPP}^1)$ is covered by two patches $U_\pm$ and therefore $\CF_\FR^{6|12}$ is covered by two patches $\tilde\CU_\pm =\FR^{4|12}\times U_\pm$. 
It is easy to see that $\diag (\CP^{3|3}\times\bar\CP^{3|3})=\hat\CU_+\cup\hat\CU_-$, where $\hat\CU_+:=\hat\CU_1$ and $\hat\CU_-:=\hat\CU_2$ are given in 
\eqref{3.1}. It follows that 
\begin{equation}\label{6.8}
\CL_\FR^{5|6}=\CU_+\cup\CU_-\quad\mbox{for}\quad\CU_\pm=\hat\CU_\pm\cap\CL_\FR^{5|6} \ .
\end{equation}
Thus, only two patches on $\CL_\FC^{5|6}$ belong to the fixed point set $\CL_\FR^{5|6}$ of the involution $\rho$. 

\medskip

\noindent
{\bf Thickening of $\CL_\FR^{5|6}$.} In section 5 we discussed the embedding
\begin{equation}\label{6.9}
S^2\subset S^2_\FC\subset \CPP^{1}\times\CPP^{1}_*\ ,
\end{equation}
where real quadric $Q^2_\FR=S^2$ parametrizes real null lines \eqref{5.7} in $\FR^{3,1}$ and complex quadric $Q^2_\FC=\CPP^{1}\times\CPP^{1}_*$ parametrizes
complex null lines \eqref{5.1}-\eqref{5.4} in $\FC^4$. The complex sphere $S^2_\FC$ is a thickening (minimal complexification) of a real-analytic sphere $S^2$ 
into a germ of complex manifold homotopically equivalent to $S^2$. Thus, we can introduce a minimal complexification  $\CL_\frC^{5|6}$ of $\CL_\FR^{5|6}$ as a
fibration over $S^2_\FC$ and obtain the double fibration 
\smallskip
\begin{equation}\label{6.10}
\begin{picture}(80,40)
\put(0.0,0.0){\makebox(0,0)[c]{$\CL_{\frC}^{5|6}$}}
\put(64.0,0.0){\makebox(0,0)[c]{$\FC^{4|12}$}}
\put(35.0,37.0){\makebox(0,0)[c]{$\CF_{\frC}^{6|12}$}}
\put(7.0,20.0){\makebox(0,0)[c]{$\pi_2$}}
\put(55.0,20.0){\makebox(0,0)[c]{$\pi_1$}}
\put(25.0,27.0){\vector(-1,-1){18}}
\put(37.0,27.0){\vector(1,-1){18}}
\end{picture}
\end{equation}
\smallskip
where $\CF_{\frC}^{6|12}:=\FC^{4|12}\times S^2_\FC$. Of course, one should remember that $S^2_\FC$ is a noncompact  complex manifold and one cannot
use a generalized Liouville theorem which states that any holomorphic function on a compact complex manifold is necessarily constant. However, \eqref{6.10} can
be used as an ``intermediate'' case between \eqref{3.15} and \eqref{6.7} according to \eqref{6.9}.

\section{Integrability of $\CN =3$ SYM on real Minkowski space}

\noindent
The twistor description of $\CN =3$ SYM theory on complexified Minkowski space $\FC^4$ is well developed (see e.g. \cite{13}-\cite{19}). The complexified theory
is described by at least six GL($N, \FC$)-valued functions $\{f_{ab}\}$  (twistor data) that satisfy the nonlinear equations \eqref{4.1} on (a domain of)
ambitwistor space $\CL_{\FC}^{5|6}$ which has dimension $(10|12$) as a real supermanifold. It is also known that SYM theory on Euclidean space 
$\FR^{4,0}\subset\FC^4$ is described by Cauchy-Riemann vector bundles over a Cauchy-Riemann submanifold $\CL_{\rm CR}^{8|12}$ of $\CL_{\FC}^{5|6}$ \cite{31}. 
Both theories (on $\FC^4$ and $\FR^{4,0}$) are not integrable since twistor data $\{f_{ab}\}$ are not free. For obtaining twistor description of $\CN =3$ SYM 
theory on Minkowski space $\FR^{3,1}$, we should restrict the complex diagram \eqref{3.15} to the real diagram \eqref{6.7} and understand what type of vector bundles
appears after restrictions of holomorphic bundles to completely real submanifolds. This is the topic of this section.

\medskip
\noindent
{\bf $\CN =3$ SYM on $\FR^{3,1}$.} We consider a trivial complex vector bundle $E=\FR^{4|12}\times\FC^N$ with the structure group U($N$) and a real analytic 
$\mathfrak{u}(N)$-valued one-form
\begin{equation}\label{7.1}
A=A_{\alpha\ald}\dd x^{\alpha\ald} + A_{\ald}^i\dd\eta^{\ald}_i + A_{\alpha i}\dd\theta^{\alpha i}
\end{equation}
with reality conditions~\cite{12}
\begin{equation}\label{7.2}
A^{\dagger}_{\alpha\dot\beta}=A_{\beta\ald}\quad\mbox{and}\quad (A_\ald^i)^\dagger =A_{\alpha i}\ .
\end{equation}
The one-form  \eqref{7.1} is  a superconnection on the bundle $E\to\FR^{4|12}$. It satisfies the real form of the equations \eqref{4.13},
\begin{equation}\label{7.3}
\{\nabla_\ald^i,\nabla_\bed^j\}+\{\nabla_\bed^i,\nabla_\ald^j\}=0~,~~~
\{\nabla_{\alpha i},\nabla_{\beta j}\}+\{\nabla_{\beta i},
\nabla_{\alpha j}\}=0~,~~~
\{\nabla_{\alpha i},\nabla^j_{\ald}\}-2\delta_i^j\nabla_{\alpha\ald}=0~,
\end{equation}
which are equivalent to the field equations of  $\CN =3$ SYM theory. Here operators $\nabla_{\alpha\ald}$, $\nabla_\ald^i$ and $\nabla_{\alpha i}$ are covariant derivatives 
given by  \eqref{4.9}-\eqref{4.11} with reality conditions.

Recall that in $\FR^{4|12}$ there is a family of real $(1|6)$-dimensional super null lines. They are defined by vector fields
\begin{equation}\label{7.4}
W=\mu^\alpha\lambda^\ald\dpar_{\alpha\ald}\ ,\quad D^i = \lambda^\ald\frac{\dpar}{\dpar\eta^\ald_i}\quad\mbox{and}\quad D_i = \overline{D^i}=
\mu^\alpha\frac{\dpar}{\dpar\theta^{\alpha i}}
\end{equation}
parametrized with the coordinates $(\lambda , \bar\lambda =\mu )$ on the sphere $\CPP^1_{\hat x}$ of null directions through the point $\hat x = (x, \theta , \eta = \bar\theta )\in \FR^{4|12}$. 
These coordinates satisfy the reality conditions \eqref{6.3}. The distribution $\CT$ generated by vector fields \eqref{7.4} forms a subbundle of the tangent bundle of  $\CF^{6|12}_\FR$.
By introducing covariant derivatives $\nabla_\CT$ along the $(1|6)$-dimensional distribution $\CT$, 
\begin{equation}\label{7.5}
\nabla_{\CT}=\{\mu_{\pm}^\alpha\lambda_{\pm}^\ald \nabla_{\alpha\ald} ,\, \lambda_{\pm}^\ald\nabla^i_{\ald} ,\, \mu_{\pm}^\alpha\nabla_{\alpha i}\}\ ,
\end{equation}
we see that equations  \eqref{7.3} are equivalent to the condition of vanishing the supercurvature
\begin{equation}\label{7.6}
F_\CT = \nabla_\CT^2=0
\end{equation}
on real super null lines in $\FR^{4|12}$.

\medskip
\noindent
{\bf Bundles $\tilde\CE$ over $\CF^{6|12}_\FR$.} Using the projection $\pi_1: \CF^{6|12}_\FR\to\FR^{4|12}$ from the diagram \eqref{6.7}, 
we pull back the bundle $E$ over $\FR^{4|12}$ to a complex vector bundle $\tilde\CE :=\pi^*_1 E$ over $\CF^{6|12}_\FR$. The bundle  
$\tilde\CE\to\CF^{6|12}_\FR$ is real analytic since it is a restriction of holomorphic vector bundle over  $\CF^{6|12}_\FC$. In accordance with 
the definition of a pull-back, a pulled back connection $\pi^*_1A$ on  $\tilde\CE$ is flat along the fibres $\CPP^1_{\hat x}$ of the projection  $\pi_1$     
and we can set its components along $\CPP^1_{\hat x}$ equal to zero. Then the pulled back covariant derivatives $\nabla_\CT$
along $\CT$ keep the form \eqref{7.5}.

In section 6 it was shown that the space $\CF^{6|12}_\FR$ is covered by two patches $\tilde \CU_\pm$. On each patch $\tilde \CU_\pm$ we have a flat relative connection
$\CA_\CT$ along the integrable distribution $\CT$ in the tangent bundle of $\CF^{6|12}_\FR$. The curvature  \eqref{7.6} of this connection is zero, i.e. $\CA_\CT$ is flat. 
Therefore, there exist U$(N)$-valued functions $\psi_\pm(x, \theta , \bar\theta , \lambda, \bar\lambda )$ on $\tilde \CU_\pm$ satisfying the equations
\begin{equation}\label{7.7}
\nabla_\CT\,\psi_\pm =0
\end{equation}
which are equations \eqref{4.9}-\eqref{4.11} with the reality conditions \eqref{6.3} and \eqref{7.2}. 

Functions $\psi_\pm$ are defined on $\tilde \CU_\pm\subset\CF^{6|12}_\FR$.
On the intersection $\tilde\CU_+\cap\tilde\CU_-$ of these patches we can introduce a U$(N)$-valued function
\begin{equation}\label{7.8}
f_{+-}=\psi_+^{-1}\psi_-
\end{equation}
and identify it with a transition function of the bundle $\tilde\CE\to\CF^{6|12}_\FR$ in another trivialization. From  \eqref{7.7}  and \eqref{7.8} it follows that
\begin{equation}\label{7.9}
\dd_\CT\,f_{+-}=0\quad\Leftrightarrow\quad W_\pm\,f_{+-}= D^i_\pm\,f_{+-}= D_i^\pm\,f_{+-}=0\ ,
\end{equation}
i.e. $f_{+-}$ is constant along the distribution $\CT$.

Note that the trivial bundle $E\to\FR^{4|12}$ is endowed with a non-trivial superconnection \eqref{7.1} subject to the equations \eqref{7.3}. The pulled-back bundle 
$\tilde\CE\to\CF^{6|12}_\FR$ is also trivial with the transition function $\tilde f_{+-}=\unit_N$ on  $\tilde\CU_+\cap\tilde\CU_-$ and it is also endowed with 
non-trivial pulled-back partial connection $\CA_\CT = \pi^*_1 A_\CT$. Matrices $\psi_\pm$ solving \eqref{7.7} for a given $\CA_\CT$ are matrix fundamental solutions, 
i.e. the columns of $\psi_\pm$ form frame fields of $\tilde\CE$ over $\tilde\CU_\pm$. In other words, $\psi_\pm$ define a trivialization $\tilde\CE$ over $\tilde\CU_\pm$.
Solutions $\psi_\pm$  of equations  \eqref{7.7} encode all information about $\CA_\CT$ and we have an equivalence of the following data
\begin{equation}\label{7.10}
\bigl(E, A\ \mbox{in}\ \eqref{7.3}\bigr)\sim \bigl(\tilde\CE ,\ \tilde f_{+-}{=}\unit_N,\ \CA_\CT\ \mbox{in}\ \eqref{7.6}\bigr)\sim 
\bigl(\tilde\CE ,\ f_{+-}{=}\psi_+^{-1}\psi_-\ \mbox{in}\ \eqref{7.9}, \ \CA_\CT{=}0\bigr)\ .
\end{equation}
Thus, after changing the trivialization of the bundle  $\tilde\CE\to\CF^{6|12}_\FR$, the partial connection  $\CA_\CT$ disappears and the information is hidden 
in a transition function \eqref{7.8}.

For a given $\CA_\CT$ one can construct $\psi_\pm$ via \eqref{7.7}. Vice versa, for given U($N$)-valued functions $\psi_\pm$ one can construct components of the 
superconnection $A$ by formulae
\begin{equation}\label{7.11}
{\psi}_+D^i_{\pm}{\psi}_+^{-1}=\psi_- D^i_{\pm}\psi^{-1}_-:=A^i_\pm=\lambda^\ald_{\pm}A_\ald^i(x,\theta,\eta)~,
\end{equation}
\begin{equation}\label{7.12}
{\psi}_+D_i^{\pm}{\psi}_+^{-1}=\psi_- D_i^{\pm}\psi^{-1}_-:=A_i^\pm=\mu^\alpha_{\pm}A_{\alpha i}(x,\theta,\eta)~,\\
\end{equation}
\begin{equation}\label{7.13}
{\psi}_+W_{\pm}{\psi}_+^{-1}
=\psi_- W_{\pm}\psi^{-1}_-:=A_{w_{(\pm)}}=\mu^\alpha_{\pm}\lambda^\ald_{\pm}A_{\alpha\ald}(x,\theta,\eta)
\end{equation}
with the reality conditions \eqref{6.3} and \eqref{7.2}. Here one uses the fact that $A_\pm^i$ is a section of the bundle $\CB (-\sfrac12, \sfrac12)\to \CPP^1$, $A_i^\pm$
is a section of the bundle $\CB (\sfrac12, \sfrac12)=\overline{\CB (-\sfrac12, \sfrac12)}$ and $A_{w_{(\pm)}}$ is a section of the bundle $\CB(0,1)\to\CPP^1$, where
complex line bundles $\CB(s,\ell)\to\CPP^1$ are described in the Appendix.

By construction, a restriction of the pulled-back bundle $\tilde\CE=\pi_1^*E$ to $\CPP^1_{\hat x}\hra\CF^{6|12}_\FR$ is analytically trivial bundle for each 
$\hat x\in\FR^{4|12}$ which is reflected in the formula \eqref{7.8}. Hence, for a given U$(N)$-valued real analytic function $f_{+-}$ on $\tilde\CU_+\cap\tilde\CU_-$
there should exist the splitting \eqref{7.8} for $f_{+-}$ restricted to $\CPP^1_{\hat x}$ and this is an analog of a Riemann-Hilbert problem. However, this is not enough
since also the space of sections of $\tilde\CE$ and $E$ should be the same. But the trivial complex rank $N$ vector bundle over  $\CPP^1$ is a direct sum of $N$ bundles
$L(0)\to\CPP^1$ of spin weight $s=0$ and it has an infinite-dimensional space of sections (spherical harmonics) discussed in the Appendix. We should demand
\begin{equation}\label{7.14}
\tilde\CE|_{\CPP^1_{\hat x}}\cong\CB(0,0)\otimes\FC^N
\end{equation}
since then
\begin{equation}\label{7.15}
\Gamma (\CPP^1_{\hat x}, \tilde\CE|_{\CPP^1_{\hat x}})=\FC^N
\end{equation}
and sections of the bundle $\tilde\CE|_{\CPP^1_{\hat x}}$ can be identified with fibres $\FC^N$ of the bundle $E\to\FR^{4|12}$ at points $\hat x\in \FR^{4|12}$.
Recall that $\CB(s,\ell)$ is a complex line bundle over $\CPP^1$ with finite-dimensional space of sections of spin weight $s$ and conformal weight $\ell$. We can interpret
\eqref{7.14} as framing of $\tilde\CE$ over ${\CPP^1_{\hat x}}$. This substitutes Liouville's theorem in the holomorphic case.

\medskip
\noindent
{\bf Bundles $\CE$ over $\CL^{5|6}_\FR$ and integrability.} We described how solutions $A$ of $\CN =3$ SYM equations on $\FR^{3,1}$ define a complex 
vector bundle $\tilde\CE$ over $\CF^{6|12}_\FR$ with real analytic transition functions $f_{+-}$ satisfying \eqref{7.9}. But \eqref{7.9} means constancy along 
tangent spaces to the real $(1|6)$-dimensional leaves of the fibration
\begin{equation}\label{7.16}
\pi_2:\quad \CF^{6|12}_\FR\to\CL^{5|6}_\FR
\end{equation}
from the diagram \eqref{6.7}. The operator $\dd_\CT$ in  \eqref{7.9} is exactly an operator of differentiation along the fibres of $\pi_2$. Hence, the bundle
$\tilde\CE$ can be identified with the pulled back bundle $\pi_2^*\CE$, where $\CE$ is a complex vector bundle over $\CL^{5|6}_\FR$ with a real analytic 
transition function $f_{+-}$. According to \eqref{7.14}, a restriction of $\CE$ to any $\CPP^1_{\hat x}\hra\CL^{5|6}_\FR$ should be the
trivial vector bundle with the conformal weight $\ell=0$,
\begin{equation}\label{7.17}
\CE|_{\CPP^1_{\hat x}}\cong\CB(0,0)\otimes\FC^N\ .
\end{equation}
By construction, this real analytic bundle $\CE$ over $\CL^{5|6}_\FR$ can be extended to a holomorphic bundle over $\CL^{5|6}_{\mathfrak C}$ and further 
to a bundle over $\CL^{5|6}_{\FC}$.

Thus, real analytic solutions of $\CN =3$ SYM model on real Minkowski space $\FR^{3,1}$ are encoded into U$(N)$-valued real analytic transition functions $f_{+-}$
which determine  vector bundles $\CE\to\CL^{5|6}_\FR$ with the property \eqref{7.17}. Solving equations \eqref{7.3} of SYM theory on $\FR^{3,1}$
is reduced to a Riemann-Hilbert-type factorization problem $f_{+-}(\hat x, \lambda , \bar\lambda )\mapsto f_{+-}=\psi_+^{-1}\psi_-$ on $\CPP^1_{\hat x}$ in 
 $\CL^{5|6}_\FR$. Hence, the $\CN=3$ SYM model on $\FR^{3,1}$ is integrable in the real analytic category.
 
\bigskip

\noindent {\bf Acknowledgements}

\noindent
This work was supported by the Deutsche Forschungsgemeinschaft grant LE~838/19.
\appendix

\renewcommand{\thesection}{\Alph{section}.}
\renewcommand{\theequation}{\thesection\arabic{equation}}

\section{Line bundles over two-sphere}
\noindent
{\bf Holomorphic line bundles $\CO(m)$.} Given the Riemann sphere $\CPP^1\cong S^2$ with standard patches $U_\pm$ and coordinates
$\lambda_\pm$ on the corresponding patches, the holomorphic line bundle $\CO(m)$ over $\CPP^1$ is defined by its transition function $f_{+-}=\lambda^m_+$,
so that complex coordinates $z_\pm$ on fibres $\FC$ over $U_\pm$ are related by formula $z_+=f_{+-} z_-=\lambda^m_+ z_-$. For $m\ge 0$, global sections of the 
bundle $\CO(m)$ are polynomials of degree $m$ in the coordinates $\lambda_\pm$ and homogeneous polynomials of degree $m$ in projective coordinates 
$\lambda_\ald$ on $\CPP^1$:
\begin{equation}\label{A.1}
p^{(m)}=f^{\ald_1...\ald_m}\lambda_{\ald_1}...\lambda_{\ald_m}\ \Rightarrow\ p^{(m)}_+=\frac{1}{\lambda^m_{\dot{1}}}p^{(m)},\ p^{(m)}_-=\frac{1}{\lambda^m_{\dot{2}}}p^{(m)}\ \mbox{and}\ p^{(m)}_+={\lambda^m_+}p^{(m)}_-\ ,
\end{equation}
where $U_+=\{\lambda_\ald\ |\ \lambda_{\dot{1}}\ne 0\}$ and $U_-=\{\lambda_\ald\ |\ \lambda_{\dot{2}}\ne 0\}$.

The line bundle $\CO (m)\to\CPP^1$ has the first Chern number $c_1=m$. The complex conjugate bundle to $\CO(m)$ is denoted by $\bar\CO(m)$ (antiholomorphic line
bundle). Its sections have transition functions $\bar\lambda^m_+: \bar z_+=\bar\lambda^m_+\bar z_-$. Hence, sections of the complex line bundle $\CO(m)\otimes\bar\CO(n)$ for $m,n\ge 0$ are homogeneous polynomials
\begin{equation}\label{A.2}
p^{(m,n)}=f^{\ald_1...\ald_m\, \alpha_1...\alpha_n}\lambda_{\ald_1}...\lambda_{\ald_m}\,\mu_{\alpha_1}...\mu_{\alpha_n}\ ,
\end{equation}
where $\mu_{\alpha}:=\overline{\lambda_{\ald}}$.

\medskip
\noindent
{\bf Hermitian line bundles $L(s)$.} The bundle $\CO(1)$ is called the hyperplane bundle over $\CPP^1$. It is associated with a principal GL$(1, \FC )$-bundle with
multiplicative group $\FC^*=\,$GL$(1,\FC )$ of nonzero complex numbers. We can reduce the structure group GL$(1,\FC )$ to the unitary subgroup U(1) and consider
the associated complex line bundle $L^{(1)}$  with the transition function $(\lambda_+/\bar\lambda_+)^{1/2}$. Its $m$-tensor power $L^{(m)}$ has the
 transition function
\begin{equation}\label{A.3}
f_{+-}=\left(\frac{\bar\lambda_+}{\lambda_+}\right)^s\ ,\quad s=-\frac{m}{2}\ ,
\end{equation}
and the first Chern class $c_1=m=-2s$. A global section $f$ of this bundle is defined by a pair of smooth complex-valued functions $f_\pm (\lambda_\pm , \bar\lambda_\pm)$
on $U_\pm$ that are related on the overlap $U_+\cap U_-$ as
\begin{equation}\label{A.4}
f_{+}=\left(\frac{\bar\lambda_+}{\lambda_+}\right)^sf_-
\end{equation}
and $f=(f_+, f_-)$ is called a function of {\it spin weight} $s$ (see e.g. \cite{34,35}). That is why we will redenote this bundle by $L(s)$ $(\equiv L^{(m)})$ to emphasize the spin
related with the action of the group SL$(2, \FC)$ on $\lambda_\ald\in \FC^2{\setminus}\{0\}$.

Equation \eqref{A.4} may be rewritten as
\begin{equation}\label{A.5}
\left(\frac{\bar\lambda_{\dot{1}}}{\lambda_{\dot{1}}}\right)^sf_{+}=\left(\frac{\bar\lambda_{\dot{2}}}{\lambda_{\dot{2}}}\right)^sf_-=:f(\lambda_{\ald}, \overline{\lambda_{\ald}})\ .
\end{equation}
Thus, \eqref{A.5} gives rise to a function $f(\lambda_\ald , \overline{\lambda_{\ald}})$ defined on $\FC^2{\setminus}\{0\}$ and satisfying 
\begin{equation}\label{A.6}
f(\vk\lambda_{\ald}, \bar\vk \overline{\lambda_{\ald}})=\left(\frac{\bar\vk}{\vk}\right)^s f(\lambda_\ald , \overline{\lambda_{\ald}})
\end{equation}
for $\vk\in\FC^*$. So, we can take \eqref{A.6} as a definition of a spin weighted function. 

Note that reduction of the groups SL$(2, \FC)$ and GL$(1, \FC)$ acting on $\FC^2{\setminus}\{0\}$ to SU(2) and U(1), respectively, is related with a positive definite
Hermitian form $h(\lambda , \bar\lambda )=\delta^{\ald\alpha }\lambda_\ald \bar\lambda_\alpha$ on $\FC^2{\setminus}\{0\}$. In local coordinates $\lambda_\pm$
on $U_\pm\subset\CPP^1$ we have
\begin{equation}\label{A.7}
h_+=\frac{1}{\lambda_{\dot{1}}\bar\lambda_1}h =1+\lambda_+\bar\lambda_+\ \mbox{on}\ U_+\ , \ \
h_-=\frac{1}{\lambda_{\dot{2}}\bar\lambda_2}h =1+\lambda_-\bar\lambda_-\ \mbox{on}\ U_-
\end{equation}
and $h_+=\lambda_+\bar\lambda_+\,h_-$ on $U_+\cap U_-$. Sections $f=(f_+,f_-)$ of the bundle $L(s)$ for $m,n\ge 0$ are
\begin{equation}\label{A.8}
f_+=\frac{1}{h^\ell_+}\, f^{\ald_1...\ald_m\,\alpha_1...\alpha_n}{\lambda^+_{\ald_1}...\lambda^+_{\ald_m}\, \bar\lambda^+_{\alpha_1}...\bar\lambda^+_{\alpha_n}}
\ \mbox{and}\
f_-=\frac{1}{h^\ell_-}\, f^{\ald_1...\ald_m\,\alpha_1...\alpha_n}{\lambda^-_{\ald_1}...\lambda^-_{\ald_m}\, \bar\lambda^-_{\alpha_1}...\bar\lambda^-_{\alpha_n}},
\end{equation}
where $m=\ell-s, n=\ell+s$ and $\lambda^\pm_\ald$ are given in \eqref{2.6}. 

Sections \eqref{A.8} form a subspace $\FC^{m+1}\otimes\bar\FC^{n+1}$ in an infinite-dimensional space of all smooth sections of the bundle $L(s)$. The
space $\FC^{m+1}\otimes\bar\FC^{n+1}$ of sections \eqref{A.8} is the same as \eqref{A.2} and it is not an irreducible representation of the group SL(2, $\FC$) or SU(2).
The irreducible representation is defined on the subspace $\CY^{(s)}_\ell\cong\FC^{2\ell+1}$ of $\FC^{\ell-s+1}\otimes\bar\FC^{\ell+s+1}$ ($\ell\ge |s|$) with symmetric
coefficients $f$'s in \eqref{A.8}:
\begin{equation}\label{A.9}
f^{(s)}_{\ell\,\pm}=\frac{1}{h^\ell_\pm}\, f^{(\ald_1...\ald_m\,\alpha_1...\alpha_n)}{\lambda^\pm_{\ald_1}...\lambda^\pm_{\ald_m}\, 
\bar\lambda^\pm_{\alpha_1}...\bar\lambda^\pm_{\alpha_n}}\in \CY^{(s)}_{\ell}\ .
\end{equation}
Thus, we have
\begin{equation}\label{A.10}
\Gamma (\CPP^1, L(s)) = \mathop{\oplus}^{\infty}_{\ell=0}\CY^{(s)}_\ell\quad \mbox{with}\quad \CY^{(s)}_\ell\cong\FC^{2\ell+1}\ ,
\end{equation}
where the spaces $\CY^{(s)}_\ell$ have an orthonormal basis $Y^{(s)}_{\ell m}$ $(m=-\ell, -\ell+1,..., \ell-1, \ell)$ of functions which are called {\it spin-s spherical harmonics} (see e.g. \cite{36}-\cite{38}). For $s=0$ they are just the usual spherical harmonics on $S^2$.

\medskip
\noindent
{\bf Complex line bundles $\CB(s,\ell)$.} In  \eqref{3.9} we introduced  holomorphic line bundles $\CO(m,n)=\mathrm{pr}_1^*\CO(m)\otimes \mathrm{pr}_2^*\CO(n)$ over $\CPP^1{\times}\CPP^1_*$ with first Chern numbers $m,n\in\FZ$. On the other hand, in \eqref{5.6} we introduced $S^2\cong\CPP^1$ as a real submanifold diag$(\CPP^1\times \overline{\CPP}^1)$ of the complex manifold $\CPP^1\times \overline{\CPP}^1$ ($=\CPP^1{\times}\CPP^1_*$). Hence, we can consider a restriction of bundles 
$\CO(m,n)$ to the diagonal that was considered in \cite{34,35}.  The restricted bundles $\CB(s,\ell)$ are smooth complex line bundles which we now describe. 

Note that sections of the bundle $\CO(m,n)$ for $m,n\ge 0$ are homogeneous polynomial $f=p^{(m,n)}$ in  \eqref{A.2} of independent projective coordinates
$\lambda_\ald$ and $\mu_\alpha$ such that
\begin{equation}\label{A.11}
f(\vk_1\lambda_\ald , \vk_2\mu_\alpha )=\vk_1^m\vk_2^n \ f(\lambda_\ald , \mu_\alpha )
\end{equation}
for complex numbers $\vk_1,\vk_2\in\FC^*$. After restriction of $\CO(m,n)$ to $\CPP^1\hra\CPP^1\times \overline{\CPP}^1$ (by putting $\mu_\alpha=\overline{\lambda_\ald}$ \cite{34,35}), we obtain a smooth complex line bundle $\CB(s,\ell)$ sections of which are homogeneous polynomials $f(\lambda_\ald , \overline{\lambda_\ald})$ such that
\begin{equation}\label{A.12}
f(\vk\lambda_\ald , \bar\vk\overline{\lambda_\ald})=\left(\frac{\bar\vk}{\vk}\right)^s(\vk\bar\vk)^\ell f(\lambda_\ald , \overline{\lambda_\ald})
\end{equation}
for $\lambda_\ald \mapsto\vk\lambda_\ald$, $\vk\in\FC^*$. Here numbers 
\begin{equation}\label{A.13}
s:=\sfrac12(n-m)\quad \mbox{and}\quad \ell :=\sfrac12(m+n)
\end{equation}
are called {\it spin weight} $s$ and {\it conformal weight} $\ell$. They are the same as in \eqref{A.8}-\eqref{A.10}. The group SL$(2,\FC )\cong\,$SO(3,1) acts on $\CPP^1$
as the group of conformal transformations. Under this action the Hermitian form $h=\delta^{\ald\alpha}\lambda_\ald\bar\lambda_\alpha$ on $\FC^2{\setminus}\{0\}$
is altered and sections \eqref{A.12} of the  bundle $\CB(s,\ell)$ carry information about the representation of the group SL$(2,\FC )$ \cite{34,35}.

The first Chern number of the line bundle $\CB(s,\ell)\to\CPP^1$ is $c_1=-2s$ and the transition function is 
\begin{equation}\label{A.14}
f_{+-}=\left(\frac{\bar\lambda_+}{\lambda_+}\right)^s(\lambda_+\bar\lambda_+)^\ell\ .
\end{equation}
Sections of the smooth complex line bundle $\CB(s,\ell)$ are polynomials
\begin{equation}\label{A.15}
p^{(s,\ell)}_{\pm}= f^{\ald_1...\ald_{\ell-s}\,\alpha_1...\alpha_{\ell+s}}{\lambda^\pm_{\ald_1}...\lambda^\pm_{\ald_{\ell-s}}\, 
\bar\lambda^\pm_{\alpha_1}...\bar\lambda^\pm_{\alpha_{\ell+s}}}
\end{equation}
with integer or half-integer $s,\ell$ given by \eqref{A.13}. In \eqref{A.15} we choose $\ell\ge |s|$. Hence the space of sections is the same as in \eqref{A.2},
\begin{equation}\label{A.16}
\Gamma (\CPP^1, \CB(s,\ell))=\FC^{\ell-s+1}\otimes \bar\FC^{\ell+s+1}\ . 
\end{equation}
Comparing \eqref{A.15} with \eqref{A.8}, we see that sections \eqref{A.15} differ from \eqref{A.8} only by factors $1/h^\ell_\pm$. This reflects the fact that $\CB(s,\ell)$ is isomorphic as topological bundle to the bundle
$L(s)$ since
\begin{equation}\label{A.17}
(\lambda_+\bar\lambda_+)^\ell = h_+^{\ell} h_-^{-\ell}=(1+\lambda_+\bar\lambda_+)^{\ell} (1+\lambda_-\bar\lambda_-)^{-\ell} \ .
\end{equation}
However, they are different from the standpoint of representation theory of the group SL$(2, \FC)$ which acts on the total space of the bundle $\CB(s,\ell)\to\CPP^1$.
Hence, the bundle $\CB(s,\ell)$ over $\CPP^1$ is a smooth counterpart of the bundle $\CO (\ell-s,\ell+s)$ over $\CPP^1{\times}\CPP^1_*$.


\end{document}